\documentclass[preprint]{aastex}
\usepackage[]{natbib}
\usepackage{apjfonts}
\usepackage{graphics}

\newcommand{\ergs}{erg$\cdot$s$^{-1}$cm$^{-2}$\AA$^{-1}$}
\newcommand{\chisq}{\ensuremath{\chi^2}\ }
\newcommand{\hbeta}{H$\beta$\ }
\newcommand{\halpha}{H$\alpha$\ }
\newcommand{\kms}{km s$^{-1}$\ }

\shorttitle{SPECTRAL MODELING OF GALAXIES} \shortauthors{LI ET
AL.}
\begin{document}

\title{Empirical Modeling of the Stellar Spectrum of Galaxies}
\author{Cheng Li, Ting-Gui Wang, Hong-Yan Zhou, Xiao-Bo Dong,
and Fu-Zhen Cheng} 
\email{leech@ustc.edu.cn} 
\affil{Center for Astrophysics, 
University of Science and Technology of China,
Hefei, 230026, China}

\begin{abstract}
An empirical method of modeling the stellar spectrum of galaxies
is proposed, based on two successive applications of Principal
Component Analysis (PCA). PCA is first applied to the newly
available stellar library STELIB, supplemented by the J, H and
K$_{s}$ magnitudes taken mainly from the 2 Micron All Sky Survey
(2MASS). Next the resultant eigen-spectra are used to fit the
observed spectra of a sample of 1016 galaxies selected from the
Sloan Digital Sky Survey Data Release One (SDSS DR1). PCA is
again applied, to the fitted spectra to construct the
eigen-spectra of galaxies with zero velocity dispersion. The first
9 galactic eigen-spectra so obtained are then used to model the
stellar spectrum of the galaxies in SDSS DR1, and synchronously to
estimate the stellar velocity dispersion, the spectral type,
the near-infrared SED, and the average reddening. Extensive tests
show that the spectra of different type galaxies can be modeled
quite accurately using these eigen-spectra. The method can yield
stellar velocity dispersion with accuracies better than 10\%, for
the spectra of typical S/N ratios in SDSS DR1.

\end{abstract}


\section{INTRODUCTION}
The observed spectrum of a galaxy is a combination of three
components: a continuum, absorption lines and emission lines. For
non-active galaxies, the continuum and the absorption line
components in the optical and near-infrared are usually dominated
by starlight, while in the mid/far-infrared they are dominated by
dust and hot gas emission. For active galaxies, the continuum is
diluted by the featureless continuum of the nucleus. For active
and non-active galaxies, the continuum is modified to varying
degrees by reddening. The emission line component is produced in H
{\sc ii} regions around hot stars or in the emission line regions
of the active nucleus. A proper decomposition of the emission line
component on one hand and the continuum plus absorption line
component (so called the stellar spectrum) on the other is the
first step toward a physical interpretation of the spectrum. For
instance, accurate measurement of emission lines is fundamental to
the identification and classification of AGN. However, the
optical spectra of many nuclei are heavily contaminated, even
dominated, by stellar absorption lines coming from the host
galaxy. Therefore, it is imperative that the underlying starlight
be properly removed before the AGN identification. As for studying
nebular emission lines and the abundances of gaseous material,
proper starlight removal is also necessary. Moreover, a
well-modeled stellar spectrum may provide direct information on
the stellar population, accurate measurement of the stellar
velocity dispersion, and hence some fundamental properties of the
host galaxy, such as spectral type, kinematics, star formation
history, etc.

Though various schemes have been proposed to subtract the
underlying starlight, the basic procedure is quite similar: a
library of absorption-line templates are built in the first place,
either from spectra of stars or from that of pure absorption-line
galaxies; secondly the templates are used to model the
non-emission-line region of the spectrum; at last the modeled
spectrum, which is taken as the integrated spectrum of the stellar
component, is removed.

When spectra of stars are employed to construct the templates, the
most common approach is stellar population synthesis.
There are two main types of population synthesis studies:
evolutionary population synthesis (Tinsley 1967; Bruzual \&
Charlot 2003) and empirical population synthesis (Faber 1972; Kong
et al. 2003). For the former, the age- and metallicity-dependent
models of stellar spectra are developed by assuming the time
evolution of a few main parameters,
and then used to reconstruct the integrated spectrum of stellar
systems. However, modern evolutionary population synthesis models
still suffer from serious uncertainties, which appear to originate
from the underlying stellar evolution theory, the
color-temperature scale of giant stars, and, for non-solar
abundances in particular, the flux libraries (Charlot, Worthey \&
Bressan 1996). The empirical population synthesis approach
reproduces the stellar spectrum
with a linear combination of the spectra of a library of stars or
star clusters. Compared to the evolutionary population synthesis,
the merit of this approach is that the result does not depend on
any assumed parameters, while the drawback is its strong
dependence on the coverage range of metallicity, spectral type and
luminosity class of the stellar library. Another drawback is that
this approach can not predict the past and future stellar
properties of galaxies.

When spectra of galaxies are used, the templates are usually
derived either from the spectrum of another galaxy with no/weak
emission lines, or from that of an off-nuclear position in the
same galaxy (Ho, Filippenko \& Sargent 1997, and references
therein). However,
the spectrum of another galaxy may not accurately reflect the
exact stellar content of the galaxy being studied, while the
off-nuclear spectrum may not be completely free of line emission.
Instead of using a single galactic spectrum as the template, which
is often chosen subjectively, Ho, Fillipenkko \& Sargent (1997)
used an objective algorithm to find the best combination of
galactic spectra to create an "effective" template. The advantage
of this modification is that the use of a large basis of input
spectra ensures a closer match to the true underlying stellar
population. Recently, Hao et al. (2002) applied Principal
Component Analysis (PCA, Deeming 1964) to several hundred pure
absorption-line galaxies and used the first few eigen-spectra as
the templates. This makes the size of the template library much
smaller because the most prominent features from the sample
concentrate into the first few eigen-spectra. The other merit is
that the best-fitting model is unique because the eigen-spectra
are orthogonal. However, both of the above methods have their
unavoidable shortcomings. Because the templates are derived from
the observed spectra of galaxies, in which the stellar velocity
dispersion is non-zero and varies from object to object, they
usually do not match the velocity dispersion of the galaxies in
question. Furthermore, the template library consists of only pure
absorption-line galaxies, which usually contain very little or no
young stellar component. As a result, the modeled spectra could
not reflect correct information on the stellar population of
emission-line galaxies, giving rise to inaccurate representative
spectra\footnote{Hao et al. also noticed this problem and
a spectrum of an A star is added during their fitting.}.

We present here a robust and efficient method of starlight removal
in the optical and the near-infrared band. Briefly, PCA is first
applied to the optical spectra of STELIB (Le Borgne et al. 2003),
a newly available stellar library in the wavelength coverage of
3500--9500\AA, and the near-infrared photometric data at J, H, and
K$_s$ bands collected from the 2 Micron All Sky Survey (2MASS;
Skrutskie et al. 1997), supplemented by the stellar library
presented by Pickles (1998) (\S 3). The stellar eigen-spectra in
the visible range are then used to model a homogeneous library of
1016 galactic spectra picked up from the Sloan Digital Sky Survey
Data Release One (SDSS DR1; Abazajian et al. 2003). PCA is again
applied to the modeled galactic spectra with zero stellar
velocity dispersion. The first 9 eigen-spectra are selected as the
final absorption-line templates (\S 4). Using these templates, the
stellar spectrum of the galaxies in SDSS DR1 can be well modeled
and the stellar velocity dispersion, the near-infrared SED, 
the spectral type and the average reddening of galaxies can be obtained
simultaneously (\S 5). Extensive tests show that the present
method is self-consistent and robust.

\section{PRINCIPAL COMPONENT ANALYSIS}
The purpose of this section is to summarize briefly the PCA method
and provide the definition of terms used in this paper.

PCA is a method used to reveal interrelations among different
variables and objects contained in a large, multivariate dataset.
Its aim is to reduce the number of dimensions in the data space,
so that the most important information can be extracted. Let the
sample being studied be a collection of $n$ objects, for each of
which there are $m$ observational variables. So one have a matrix
${\bf X}=\{x_ {ij}\}_{n\times m}$ ($i=1,\cdots,n$ and
$j=1,\cdots,m$), each row vector of which corresponds to the
different variables of a given object, while each column vector,
to the same variable of the various objects. PCA proceeds from the
given matrix ${\bf X}$ and yields $m$ new variables, the principal
components (PCs), which are mutually independent and generally,
the first $m^{\prime}$ ($m^{\prime} \ll m$) PCs contain a majority
of the information of the data. Each PC is a linear combination of
the original $m$ variables; the corresponding coefficient vector
is called eigenvector. Furthermore, via the eigenvectors, the
original $m$ variables of each object are projected onto the PCs
to yield the new variables of this object. For example, the $j$-th
PC of $i$-th object is given by
\begin{equation}
pc_{ij} = {\rm\mathbf{e_j}}\cdot {\rm\mathbf{x_{i}}} = e_{j1}
x_{i1} + \cdots + e_{jk} x_{ik} + \cdots + e_{jm} x_{im}\,,
\end{equation}
where $\rm\mathbf{x_{i}}$ is the $i$-th row vector of
$\mathbf{X}$, and $\rm\mathbf{e_j}$ is the eigenvector of the
$j$-th principal component.

The fundamental principles of PCA could be understood as follows.
The contribution of PCs to the variance of the original dataset is
a measure of the amount of original information contained in the
PCs. Thus, the destination of PCA method is to seek the set of
eigenvectors that give rise to PCs with the maximum variance.
According to the theory of statistics, the expected eigenvectors
are essentially the orthogonal eigenvectors of the covariance
matrix ${\bf C} =\{c_{jk}\}_{m\times m}$, where
\begin{equation}
c_{jk}=\frac{1}{n-1}\sum_{i=1}^n(x_{ij}-\bar{x_j})(x_{ik}-\bar{x_k})\,
 ;\hspace{6mm}  1\leq j\,, k\leq m\,.
\end{equation}
Accordingly, the covariance matrix ${\bf C}$ is first constructed.
Then the determinant equation $|{\bf C}-l{\bf I}|=0$ is solved to
find the $m$ eigenvalues $\{l_j\}$ ($l_1\geq l_2\geq
l_3\geq...\geq l_m\geq 0$), where ${\bf I}$ is the unit $m\times
m$ matrix. After this, by solving the equation $({\bf C}-l_j{\bf
I}){\bf e_j}=0$, the eigenvectors ${\bf e_j}$ are obtained. Since
there are $m$ eigenvalues, there will be at most $m$ eigenvectors.
The PCs are ordered by decreasing the eigenvalues, which are
generally used to characterize the contribution of the
corresponding principal components to the original information.
The quantity $l_j/\sum\limits_{i=1}^{m}l_i$ is called the
fractional or relative contribution of the $j$-th principal
component, and
$\sum\limits_{j=1}^{k}l_j/\sum\limits_{j=1}^{m}l_j$, the
cumulative contribution of the first $k$ principal components. In
general, it is not necessary to find all the $m$ principal
components: most of the information is contained in the first few,
with the first one having the lion's share.

Much has been written about the use of PCA in studies of the
multivariate distribution of astronomical data (Connolly et al.
1995, and reference therein). In most of earlier studies, the
variables of the original dataset are observed (generally
normalized) fluxes at $m$ wavelength channels of $n$ celestial
bodies (objects), giving a matrix of ${\bf X}=\{x_ {ij}\}_{n\times
m}$; the resultant $m$ eigenvectors with $m$ wavelength channels
for each eigenvector are so called eigen-spectra, and used for
further applications: spectral classification, modeling of
stellar spectra, etc. In this paper, we perform PCA in a slightly
different way: the input data matrix is transposed to be ${\bf
X}=\{x_ {ij}\}_{m\times n}$ before being analyzed, i.e. the
variables now become the fluxes of various celestial bodies at
given wavelength channel, and the objects are thus all the
wavelength channels. In this case, PCA carries out an $n\times n$
matrix of eigenvectors, corresponding at most $n$ principal
components. Thus, the projection of the $n$ celestial bodies onto
the principal components gives rise to $n$ new spectra, which
denotes $n$ "new" celestial bodies, called "eigen-star" or
"eigen-galaxy". We then use these spectra as our templates for
modeling.

The main difference between the above methods is easy be to
understood. In the former case, the eigenvectors (eigen-spectra)
correlate strongly with the prominent features present in the
spectra of celestial bodies; thus the projection of a celestial
spectrum onto a principal component gives a measure of the
relative contribution of this celestial spectrum to the
corresponding eigen-spectrum. However, in the latter, it is the
eigenvectors that represent the contributions of the celestial
bodies to the eien-stars or eigen-galaxies, while the spectra of
the eigen-stars or eigen-galaxies contain various spectral
features presented in original celestial spectra. We argue that,
in some sense, the matrix-transposed method is much more effective
and convenient. It is well known that a significant drawback of
implementing PCA on large or very high-dimensional data sets is
the required computation time (Madgwick et al. 2003). For $n$
spectra, each with $m$ wavelengths, this requires $O(nm^2)$
operations. Therefore, if the data matrix is transposed, the
operation will become $O(mn^2)$, which is much smaller in the case
of $m\gg n$. Given that each spectrum contains $O(10^3)$
wavelength channels and the number of spectra is $O(10^2)$, the
latter method would reduce the computing expense by a magnitude.
It goes without saying for analyzing spectra of higher resolution.

For convenience, the spectra of eigen-stars or eigen-galaxies, i.e
the projection of the spectra of stars or galaxies onto principal
components, are still called eigen-spectra in the following text,
though its meaning is different from that in general case.

\section{DERIVATION OF STELLAR EIGEN-SPECTRA}
The stellar library is the foundation of applying the present
method. It is conceivable that the quality of fits to observed
galaxy spectra seriously hinges on the resolution of the stellar
spectra, as well as the coverage of spectral type, luminosity
class, chemical abundance, and wavelength range.

Theoretical stellar spectra are often preferred for spectral
modeling, because of their uniformity and generally more
extensive coverage (e.g Kurucz 1992; Lejeune 1997,1998; Westera et
al. 2002). However, at the time of writing, no such library at the
spectral resolution similar to/higher than that of SDSS spectra is
available. Furthermore, these libraries may be biased in color and
line strength, as many minor contributors to stellar opacity and
to the emergent spectrum usually cannot all be included for the
full spectral range because of computational constraints (Pickles
1998).

Several observed stellar libraries have been published, covering
the ultraviolet (Heck et al. 1984; Robert et al. 1993; Walborn et
al. 1995), optical(Gunn \& Stryker 1983; Jacoby, Hunter \&
Christian 1984; Pickles 1985; Kiehling 1987), and
near-infrared(Danks \& Dennefeld 1994; Serote, Boisson \& Joly
1996) wavelength ranges. They are usually obtained with different
instrumentation, at different resolution, spectral sampling and
for different purposes (Pickles 1998). The most common used
stellar library is HILIB, presented by Pickles (1998), which
consists of 131 flux-calibrated spectra with complete wavelength
range from 1150 to 10620 \AA\ in steps of 5 \AA. However, the
metallicity coverage of this library is limited to be near solar
abundance, and the spectral resolution is relatively lower for
accurately spectral modeling.

High resolution libraries spanning a wide range in metallicity,
spectral type and luminosity class have only recently become
available. The STELIB library (Le Borgne et al. 2003), a new
spectroscopic stellar library, consists of an homogeneous library
of over 250 stellar spectra covering the wavelength range from
3200 to 9500 \AA, with a resolution of $\lesssim$ 3 \AA\ (1\AA\
sampling) and a signal-to-noise ratio of $\sim$ 50. The library
includes stars of a range of metallicity from 0.05 to 2.5 times
solar, a range of spectral type from O5 to M6 and luminosity class
from I to V. Because of its wide coverage of spectral resolution
and spectral type, this library represents a substantial
improvement over previous stellar libraries (Le Borgne et al.
2003).

Although the visible region is always favored, the near-infrared
is equally important because it reveals stars that remain hidden
in the visible by interstellar dust (Combes et al. 2002). To
extend the library into near-infrared band, we extract the
infrared photometric data from 2MASS, which has mapped the full
sky at three near-infrared wavelengths with 10 $\sigma $
sensitivity limits of J=15.8, H=15.1, and $K_{s}$=14.3 mag.

We therefore use STELIB as our star library. The optical spectra
in STELIB and the corresponding near-infrared photometric data in
2MASS supplemented with HILIB are incorporated to form a star
library of highest quality to date.

\subsection{Optical Data}
\begin{figure}
\epsscale{0.6} \plotone{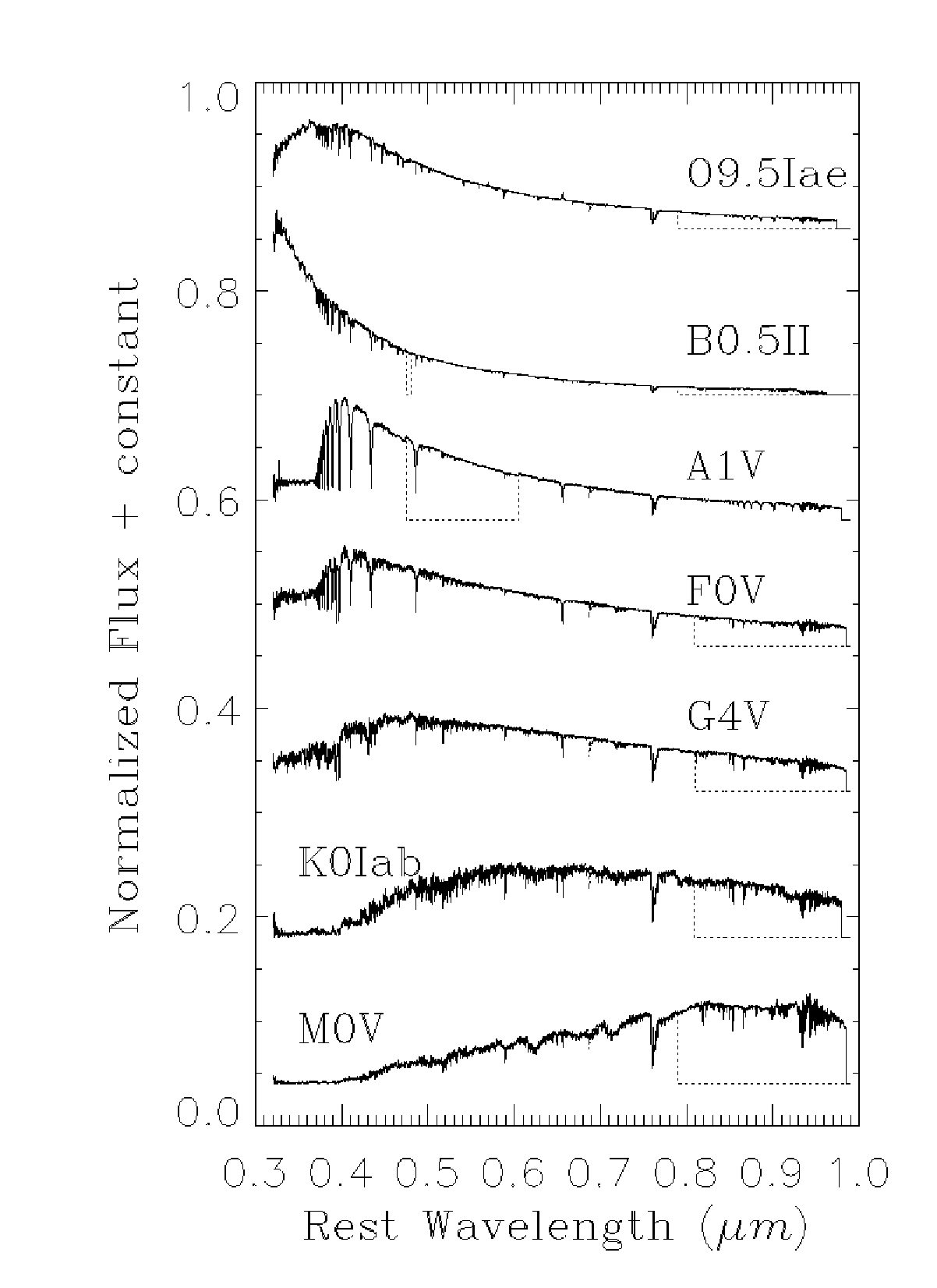}
\caption{ Examples of the filled
spectra. The dashed lines represent the initial spectra in STELIB;
the solid are the corresponding filled ones. Arbitrary constants
were added to the scaled spectra for clarity. } \label{f1}
\end{figure}

The current public version of the STELIB
library\footnote{http://webast.ast.obs-mip.fr/stelib} contains 255
optical spectra, which had been corrected for interstellar
extinction and radial velocity.
However, the data are missing in some limited wavelength range for
nearly half of the stars. We use similar stellar spectra either
from STELIB itself or from SDSS DR1 to fill the gaps of these
spectra. Some examples of the result of this procedure are shown
in Fig. \ref{f1} and at last 204 spectra are yielded.

\begin{figure}
\epsscale{0.6} \plotone{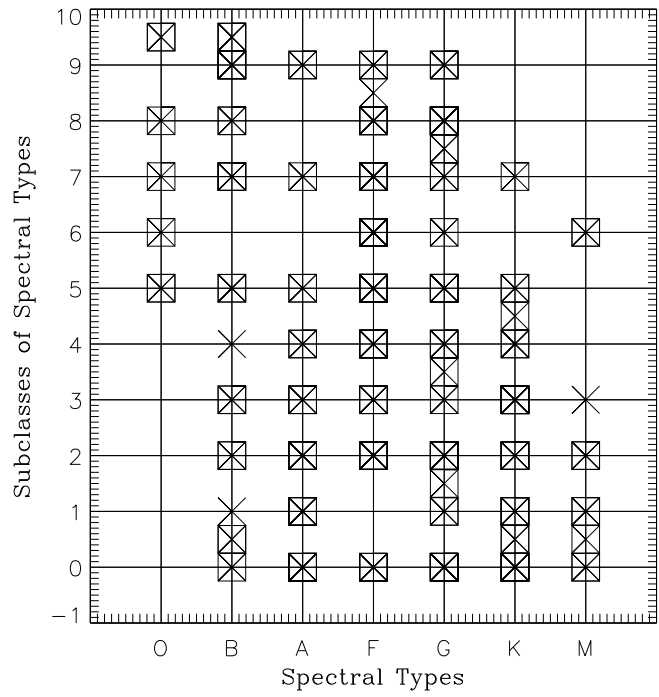}
\caption{ Distribution of the spectral
types of the STELIB library (cross) and that of the 204 stars used
in this paper (squared). Note that, out of the total 255 stars in STELIB,
Feige110 (dwarf A), LTT4364 (dwarf Carbon), and
6 WC- or WN-type stars are not plotted in this figure. }
\label{f2}
\end{figure}

Although about 50 spectra that could not be satisfactorily filled
were ignored, the resultant stellar library is still homogeneous
enough. The distributions of spectral types of the initial 255
stars and that of the final 204 stars are showed in Fig. \ref{f2}.
It is apparent that the coverage of spectral type of the resultant
library is almost as good as that of the original.

There is no denying that, over previous stellar libraries,
substantially improved as STELIB is, it still leaves much to be
desired. As can be seen from Fig. \ref{f2}, there are obviously
two blank regions: early O, and late K, M and L types. The absence
of these types of stars may have some influence on modeling the
stellar spectrum of galaxies, especially for those mainly
consisting of late-type stars.

\subsection{Near-Infrared Data}
We search for counterparts of the 204 stars in the 2MASS Pointed
Source Catalog. Out of them, 182 have been detected in 2MASS. The
magnitudes of these stars at J (1.235$\pm$0.006 $\mu m$), H
(1.662$\pm$0.009 $\mu m$) and K$_{s}$ (2.159$\pm$0.011 $\mu m$)
bands are then transferred to fluxes $f_J$, $f_H$ and $f_K$ in
unit of \ergs, according to the following
formulas (see Mannucci et al. 2001),
\begin{equation}
\begin{array}{lll}
f_J = 3.129\times 10^{-0.4(J+25.0)} & , &
\Delta f_J=(0.0175-0.4\Delta J)\cdot f_J \\
f_H = 1.133\times 10^{-0.4(H+25.0)} & , &
\Delta f_H=(0.0195-0.4\Delta H)\cdot f_H \\
f_K = 4.283\times 10^{-0.4(K+27.5)} & , & \Delta
f_K=(0.0188-0.4\Delta K)\cdot f_K
\end{array}
\end{equation}

As for the remaining 22 stars, the three near-infrared fluxes are
derived from the similar spectra in HILIB. For each of the 22
optical spectra in STELIB, a cross-correlation method is performed
to find the best-matched spectrum among the 131 spectra in HILIB,
and the corresponding fluxes at J, H, and K$_{s}$ bands are
accepted to be the real values of the NIR data of this star.

\subsection{PCA of Stellar Library}
Before performing PCA, the 204 spectra in visible range are
trimmed to the common wavelength range of 3500 -- 9500\AA, and,
each spectrum including the fluxes at J, H, and K$_{s}$ bands, is
normalized to unit, $\sqrt{\sum_\lambda f^2_\lambda}=1$ (Connolly
et al. 1995), where $f_\lambda$ is the flux at wavelength
$\lambda$. It is universally acknowledged that the set of
orthogonal eigenvectors resulting from PCA is affected by the
scaling of the data because the scaling may affect lines and
continuum differently (Sodr$\rm\acute{e}$ \& Cuevas 1997). Since
the data here are spectra of stars or modeled stellar spectra of
galaxies (see \S 4) which are non-emission-line, the method of
normalization does not affect the results of PCA. To verify this
argument, we have repeated the analysis with other methods of
normalization, e.g $f_{5500}=1$ (Kennicutt 1992) and $\sum_\lambda
f_\lambda=1$ (Sodr$\rm\acute{e}$ \& Cuevas 1997), and found that
the results are almost independent of the adopted normalization.

\begin{figure}
\epsscale{0.6} \plotone{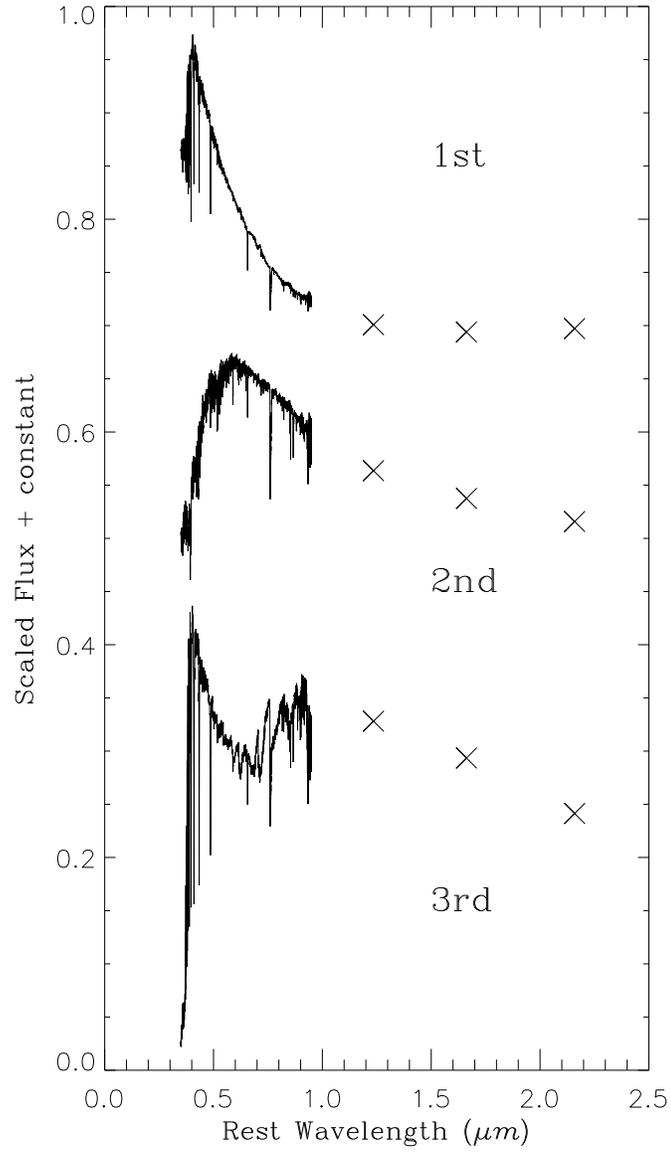}
\caption{The first three eigen-spectra of the stellar library.
The solid lines represent the spectra in visible range, while
the cross points are the J, H, and K$_s$ fluxes.
Arbitrary constants are added to the scaled spectra for clarity. } \label{f3}
\end{figure}

\begin{figure}
\epsscale{0.6} \plotone{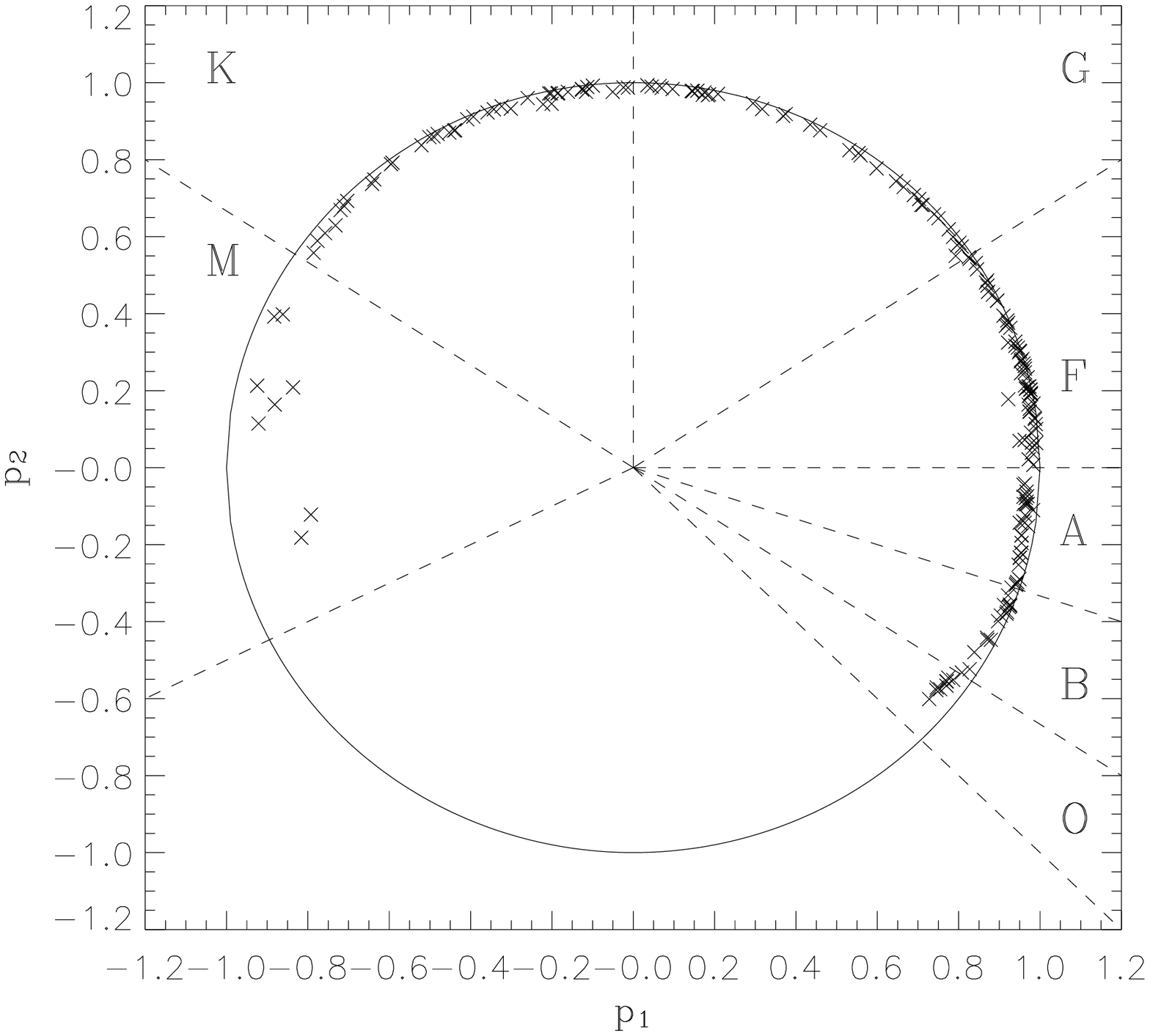} \plotone{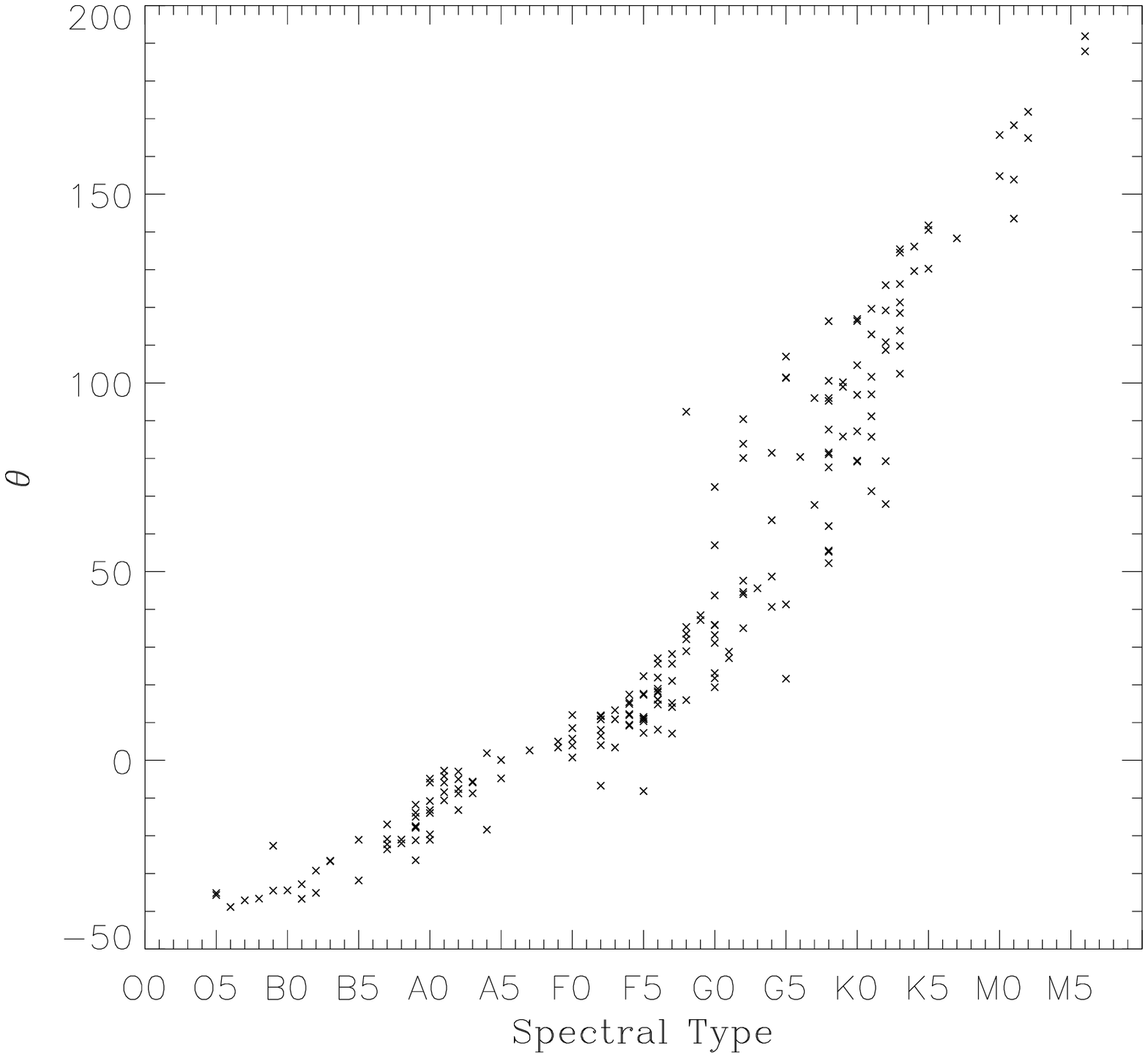}
\caption{
{\it Left}: 
Relative contributions of the 204 stars to the first two eigen-stars.
{\it Right}:
The position angles $\theta$ of the 204 stars in the
$p_1$ {\it vs}\rm\ $p_2$ plane (left panel), 
as a function of their spectral types.
$\theta$ is measured anticlockwise from the dashed horizontal line
in the left panel.}
\label{f4}
\end{figure}

The 204 scaled spectra are then analyzed by PCA after the data
matrix being transposed (see \S 2). The fractions of contribution
to variance by the first three eigen-spectra (Fig. \ref{f3}) are
61.7\%, 34.4\% and 1.7\%. In Fig. \ref{f4} (left panel), 
the 204 stars are
plotted in the plane of $p_1\ vs\ p_2$, where $p_1$, $p_2$ are the
relative contributions by each star to the first 2 eigen-spectra.
It is remarkable that most of the stars distribute on the circle
with $r=\sqrt{p_1^2+p_2^2}\simeq 1$. Only few very late- and
early-type stars deviate from this unit circle scattering well
within $r=1$. This result is understandable because the cumulative
contribution to variance by the first two eigen-stars is $96.1\%$
and this indicates that they contain most of the information of
the original 204 stars. For those stars scattering far from the
unit circle, the contribution of other eigen-spectra is important.
In the right panel of Fig. \ref{f4}, 
the position angles of the 204 stars in the left panel
are plotted, against their spectral types. As can be seen
in Fig. \ref{f4}, the 204 stars are well separated on
the circle and ordered by spectral types from early to late.
Therefore, PCA method can provide a new way of spectroscopic
classification of stars.

These eigen-spectra of the stellar library will be used as 
absorption-line templates to model the spectra of galaxies in 
the next step. As it has been pointed out in the last section,
one of the advantage of the PCA method is that only a few of the
first eigen-spectra are enough for spectral modeling. In order
to determine the number of significant eigen-stars, we estimate
the expected level of the variance caused by the noise in the
spectra as
$\sum_{i=1}^{204}\sum_{j=1}^{m_{p}}\sigma_{ij}^2/
\sum_{i=1}^{204}\sum_{j=1}^{m_{p}}(f_{ij}-\bar{f_{i}})^2$, where
$m_p$  is the number of wavelength channels, $f_{ij}$ and
$\sigma_{ij}$ the flux and its error of the $j$th  wavelength
channel of the $i$th star, and $\bar{f_{i}}$ the average flux
of the $i$th star. The errors $\sigma_{ij}$ are
determined by assuming a signal-to-noise ratio of $S/N=50$,
which is the typical value for STELIB.
This estimation gives rise to a significance of
0.2\% and the number of siginificant eigen-stars of 24,
indicating that the first 24 eigen-stars together contribute nearly all
the useful information of the stellar library.

\section{DERIVATION OF GALACTIC EIGEN-SPECTRA}
                                                                                
\subsection{Selection of Galactic Templates}\label{samsel}
A full spectral type coverage is crucial to a library of
galaxies, as it is to that of stars. SDSS is to date the most
ambitious imaging and spectroscopic survey, and will eventually
cover a quarter of the sky (York et al. 2000). The large
coverage of area and
moderately deep survey limit of the SDSS make it be very
propitious to construct a library of template galaxies with full
spectral types.

In the first step, all the low redshift ($z<0.2$) and high
signal-to-noise ratio ($(S/N)_g>30$ or $(S/N)_r >40$ or
$(S/N)_i>40$) objects in the SDSS DR1 spectroscopically classified
as galaxies by the SDSS pipeline are selected as candidates of
the template galaxies. The resultant 7098 spectra are transformed
to the rest frame using the redshift provided by the SDSS
spectroscopic pipeline.

The 7098 spectra are then fitted with the first 24 eigen-spectra of
the star library obtained in the last section. The eigen-spectra
are broadened to velocity dispersions from 0 to 600
\kms using Gaussian kernel.
To avoid the effect of emission lines, the central $\sim$5\AA\ of
the emission lines (e.g, Balmer system, forbidden
lines) are excluded from the fit. The best-fitting model of each
spectrum is therefore derived through the \chisq minimization by
taking into account of flux uncertainties provided by the SDSS pipeline.
Given the 24 star eigen-spectra as the input templates, our program
solves for the systemic velocity, the line-broadening function
 and the relative contributions of the various templates. The best-fitting
model, in general, is a good solution for the absorption-line
spectrum of the stellar component.
                                                                                
The large size of the galaxy sample provides us with an extensive
collection of modeled spectra spanning a full range of spectral
type. However, the distribution of the spectral types in the 
galaxy library is not uniform.
To get a uniform library, we re-select candidates on a
set of color-color diagrams. Each modeled spectrum is cut into
18 pass-bands with widths of $\sim 200$\AA\ , and an synthesized
magnitude of each band is obtained via
\begin{equation}
\begin{array}{ll}
m_i = -2.5 \lg \int f_\lambda^i d\lambda, & i=1,2,...,18,
\end{array}
\end{equation}
where $f_\lambda^i$ is the spectrum in the $i$th waveband.
The color-color diagrams should make use of all the 18
magnitudes, whereas it does not mean that we need to
make use of all the possible color-color diagrams.
We assign the 18 magnitudes into 6 groups,
each of which have 3 magnitudes.
For each group, the 3 magnitudes give rise to 2 colors,
$c_1$ and $c_2$,
\begin{equation}
\begin{array}{lll}
c_{i1} = m_i - m_{i+6}, & c_{i_2} = m_{i+6} - m_{i+12}, &
i=1,2,...,6,
\end{array}
\end{equation}
where $c_{i1}$ and $c_{i2}$ are the two colors in the $i$th
group. In this way, all the 18 magnitudes but only 6 color-color
diagrams are used.
                                                                                
\begin{figure}
\epsscale{0.6} \plotone{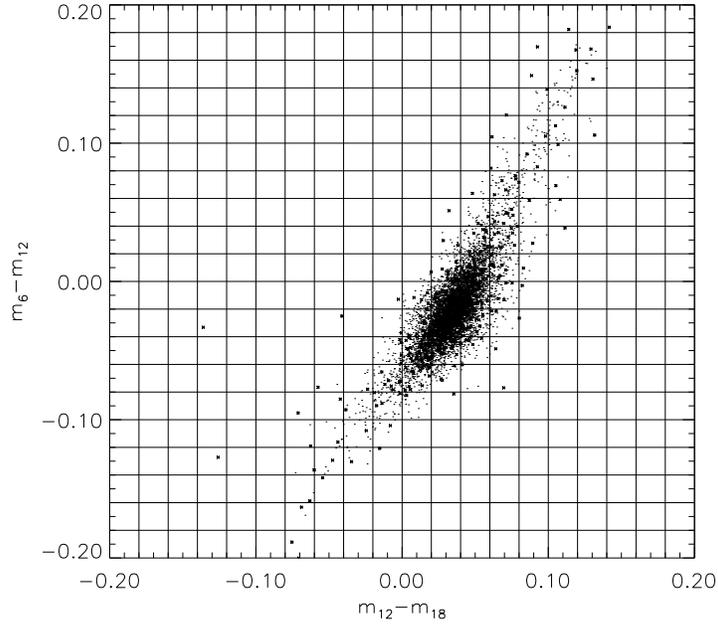}
\caption{An example of selecting galaxy sample on color-color diagrams.
$m_6$, $m_{12}$, and $m_{18}$ are the magnitudes derived from the
modeled spectrum of galaxies (see the text for details).
{\it dot}: 7098 galaxies of low redshift and high spectral S/N ratio
in SDSS DR1. {\it cross}: the galaxies selected on this diagram.}
\label{f5}
\end{figure}

The candidates are then selected on the 6 color-color diagrams.
To the end, each diagram is partitioned into square meshes with
the size of 0.02 magnitudes. The mesh size is so chosen as to have
nearly a full coverage of spectral type for the galaxy sample.
The galaxies located within each mesh are ordered by decreasing
the signal-to-noise ratios of their spectra, and the first 10\%
are picked up. The galaxy with the highest spectral
signal-to-noise ratio is alone selected if there are less than 10
galaxies in a mesh. 
As an example, Fig. \ref{f5} shows the selection procedure on the
$m_6-m_{12}$ {\it vs} $m_{12}-m_{18}$ diagram.
1126 unique galaxies are obtained in this stage. Then each spectrum 
is examined by eye and 110 objects are rejected because of either the 
presence of bad wavelength channels in the spectrum or the contamination 
of nuclear activity. At last 1016 galaxies are chosen as our galaxy 
templates.

\subsection{Iterative Spectral Modeling Using Stellar Templates}
\label{modeling}
The procedure of spectral modeling for the 7098 template
candidates is rather rough, although it is sufficient for sample
selection. In fact, the following issues that might affect the fit
should be carefully addressed.

First, masking precisely the emission line region is vital to the
measurement of the profile of absorption lines, which are sometimes
filled partially with one or more emission lines. A narrower masked
range may include wings of the emission line, while a wider one will
exclude also useful information of the absorption line profile.
Besides the width and equivalent width of emission lines differ
from line to line, and from object to object. A specific masked region
for each line in each object is required for such subtle treatment.
In this paper, we use the measured emission parameters to create the mask
region for each emission line of each object.

Second, galaxies often suffer from intrinsic reddening to certain
extend. In this paper, we estimate the intrinsic extinction of
galaxies in a way similar to that commonly used by population
synthesis, i.e., a single extinction for the whole galaxy. 
Certainly this is only a zero order solution, and by no means a
satisfactory one.  During the modeling, the program searches a range of
color excess $E(B-V)$ to find the most plausible value 
by assuming an extinction curve of Calzetti et al. (2000).

Finally, bad pixels in the SDSS spectrum, flagged by the SDSS pipeline,
or in the stellar templates are also masked from fitting.

\begin{figure}
\epsscale{0.6} \plotone{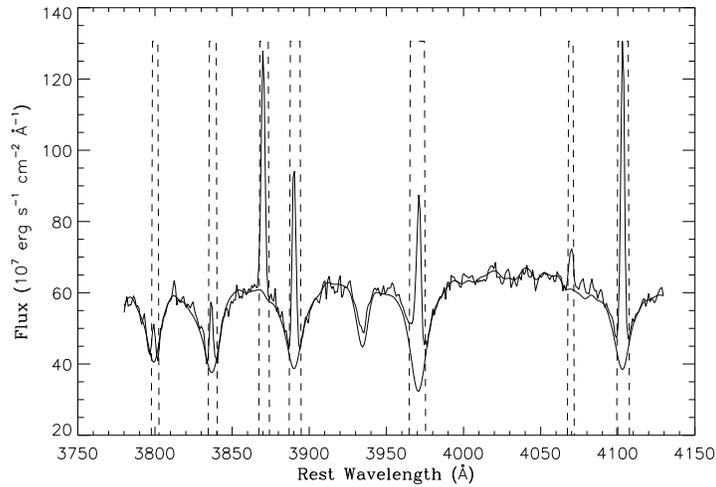}
\caption{An example of masked regions of emission lines during the spectral modeling.
The thine line is the observed spectrum of SDSS J1308+0020, and
the thick is the best-fitting spectrum. The dashed lines denote the masked regions.}
\label{f6}
\end{figure}

We use an iterative procedure to re-model the 1016 galaxies by taking
care of above issues. Each spectrum is modeled at least three times.
Initially, the same procedure as described in \S \ref{samsel} is
performed. Then the modeled spectrum is subtracted from the
observed one, and emission lines are fitted with Gaussian functions
(see Dong et al. 2004 for details). Because emission lines and absorption
lines are coupled, this procedure must be performed iteratively.  An average
reddening is added to model. Detailed parameters are listed below:
\begin{itemize}
\item The range of $E(B-V)$ is set to be from 0 to 2 with a step size of 0.01.
\item Pixels with emission line flux above 3 $\sigma$, the flux uncertainty
of SDSS at that pixel, are masked.
See Fig. \ref{f6} for an example of this procedure.
\item Pixels in the wavelength ranges from 6800 to 7100\AA\ 
and from 7500 to 7700\AA\ in the source rest frame are
masked due to the atmosphere absorption in the original stellar library.
\item Pixels within 100\AA\ of the left-end and 200\AA\ of the right-end 
of each spectrum are masked to avoid possible calibration problem.
\end{itemize}

\subsection{PCA of Galactic Library}
Galaxy template spectra with effectively zero velocity dispersion
are obtained according to the procedure described in \S4.2.
Given a set of expansion coefficients, the modeled spectrum with
zero velocity dispersion can be obtained via
\begin{equation}\label{fit}
f_\lambda = \sum_{i=1}^{24}a_ie_{i\lambda},
\end{equation}
where $e_{i\lambda}$ is the $i^{th}$ eigen-star and $a_i$ the
best-fitting coefficients.

\begin{figure}
\epsscale{0.6} \plotone{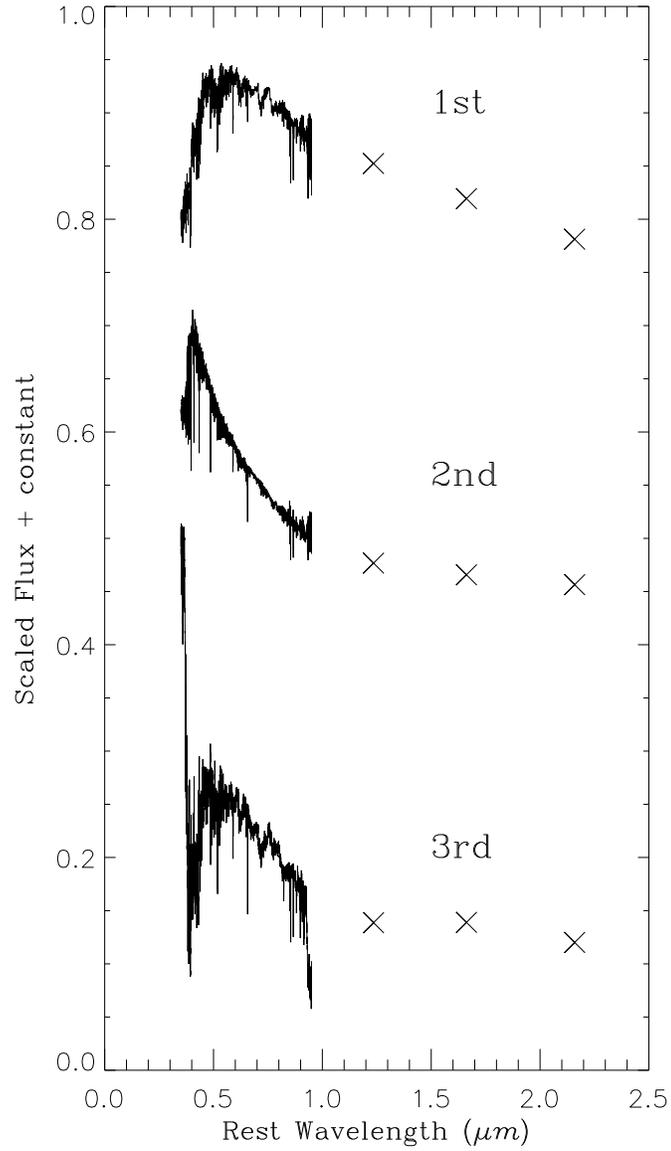}
\caption{The spectra of the first three eigen-galaxies. 
The solid lines represent the spectra in visible range, while
the cross points are the J, H, and K$_s$ fluxes.
Arbitrary constants are added to the scaled spectra
for clarity. } \label{f7}
\end{figure}

\begin{figure}
\epsscale{0.6} \plotone{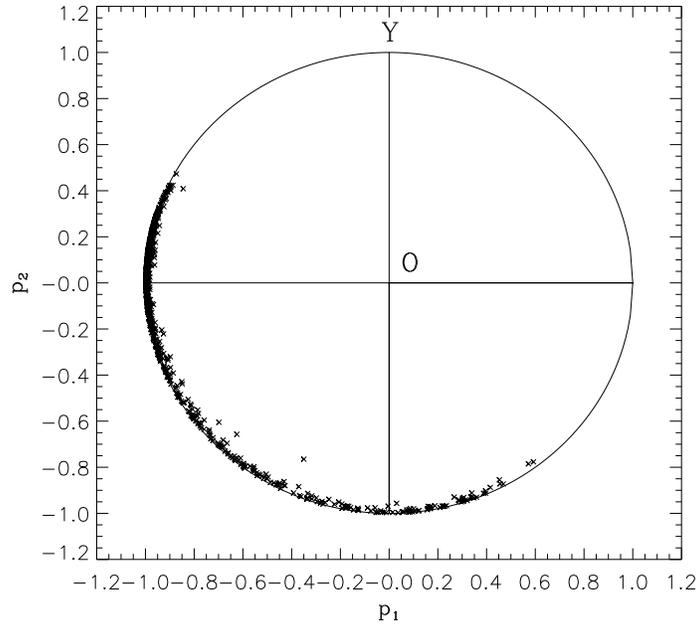}
\caption{Relative contributions of the 1016 galaxy templates
to the first two eigen-galaxies.} \label{f8}
\end{figure}

Then the 1016 zero-velocity-dispersion spectra are analyzed by
PCA, in the same way as in \S3.3. 
Because the modeled spectra are constructed using the first 24
star eigen-spectra, only the first 24 eigen-galaxies are non-zero.
The fractional contributions of the first 3 eigen-galaxies to variance 
are 81.4\%, 17.0\% and 0.4\%. Fig. \ref{f7} shows the spectra of these
3 eigen-galaxies, including their fluxes at J, H and K$_{s}$
bands. As in the case for stellar library, the variance is so dominated 
by the contributions of the first two eigen-galaxies that most of galaxies 
are distributed around a circle of radius $r=\sqrt{{p_1}^2+{p_2}^2}=1$ on 
$p_1$ vs $p_2$ plane (see Fig. \ref{f8}).

\begin{figure}
\epsscale{0.6} \plotone{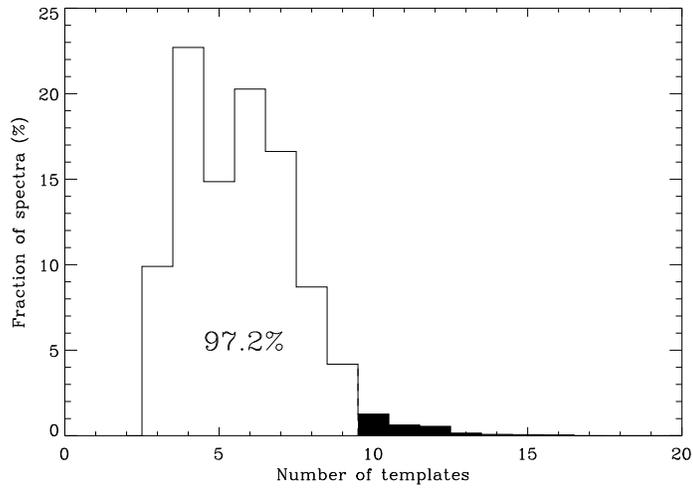}
\caption{F-test to choose the number of significant eigen-spectra of
galaxies.} \label{f9}
\end{figure}

The number of eigen-galaxies that are required to fit 
the observed galaxies is determined as follows. Initially, the observed 
spectra are modeled using the first 3 eigen-galaxies. By adding successively 
the next eigen-galaxy to the model, we calculate the significance of the 
improvement to the fit using F-test:
\begin{eqnarray}
\alpha_F & = & \int_F^\infty dF p (F|\Delta P, N-P_1) \\
         & = & I_{\frac{N-P_1}{N-P_1+\Delta P \cdot F}}
\big(\frac{N-P_1}{2},\frac{\Delta P}{2}\big), \\
\end{eqnarray}
where, the $F$-statistics $ F=\frac{\Delta \chi^2 / \Delta P} 
{\chi_1^2 / (N-P_1)}$, 
$P_1$ and $\Delta P=1$ are the numbers of thawed parameters of the previous 
 model and of the additional freely varying parameters in the current 
model, $I$ the incomplete beta function. 
Adopting a critical significance of $\alpha_F=0.05$, we find that more than 
97\% galaxies can
be well-modeled using the first 9 eigen-spectra (see Fig. \ref{f9}),
which are then chosen as our final absorption-line templates.

\section{APPLICATIONS AND TESTS}

\subsection{Modeling the Spectra in SDSS DR1}

\begin{figure}
\epsscale{0.6} \plotone{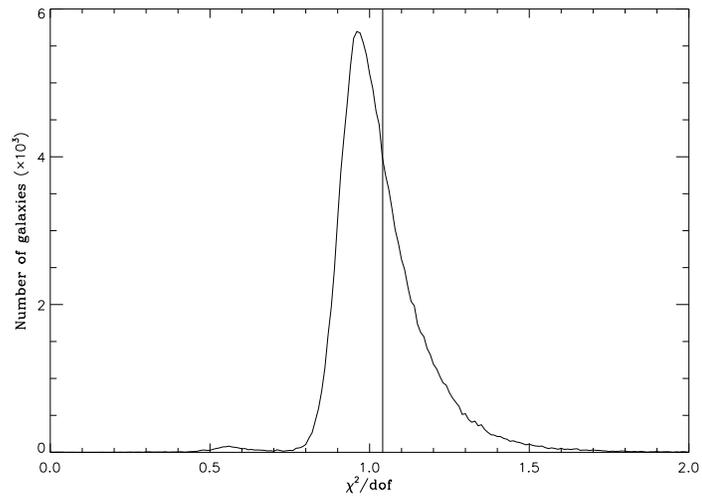}
\caption{
Distribution of reduced \chisq for spectral modeling of all
galaxies in SDSS DR1. The vertical line marks the mean value
of $\chi^2/dof$=1.04.
}\label{f10}
\end{figure}

The spectra of all galaxies in SDSS DR1 ($\sim 1.4\times 10^5$
spectra) are fitted with the 9 galactic eigen-spectra with
iterative rejection of emission lines and bad pixels (see \S \ref{modeling}). 
On a personal computer with a CPU of main
frequency 2.8 GHz, this procedure takes $\sim$20 hours.
The reduced \chisq, namely $\chi^2_\nu$ (=$\chi^2/dof$, where $dof$ is degree of freedom), 
for 98.8\% spectra are less than 1.5,
with the peak of the $\chi^2_\nu$ distribution around 0.96,
and the mean $\chi^2_\nu\sim$1.04,
indicating that the fits are quite good (see Fig. \ref{f10}).
Fig. \ref{f13} shows the
contour of the number density of DR1 galaxies on the $p_1^\prime$
{\it vs} $p_2^\prime$
plane, where $p_1^\prime$ and $p_2^\prime$ are the fractional contributions of
the first 2 galactic eigen-spectra to the modeled spectra.
On this diagram, 98.6\% galaxies in SDSS DR1 locate in the ring with
a radius $\sqrt{{p_1^\prime}^2+{p_2^\prime}^2}$ between 0.9 and 1.

\begin{figure}
\epsscale{0.35}
\plotone{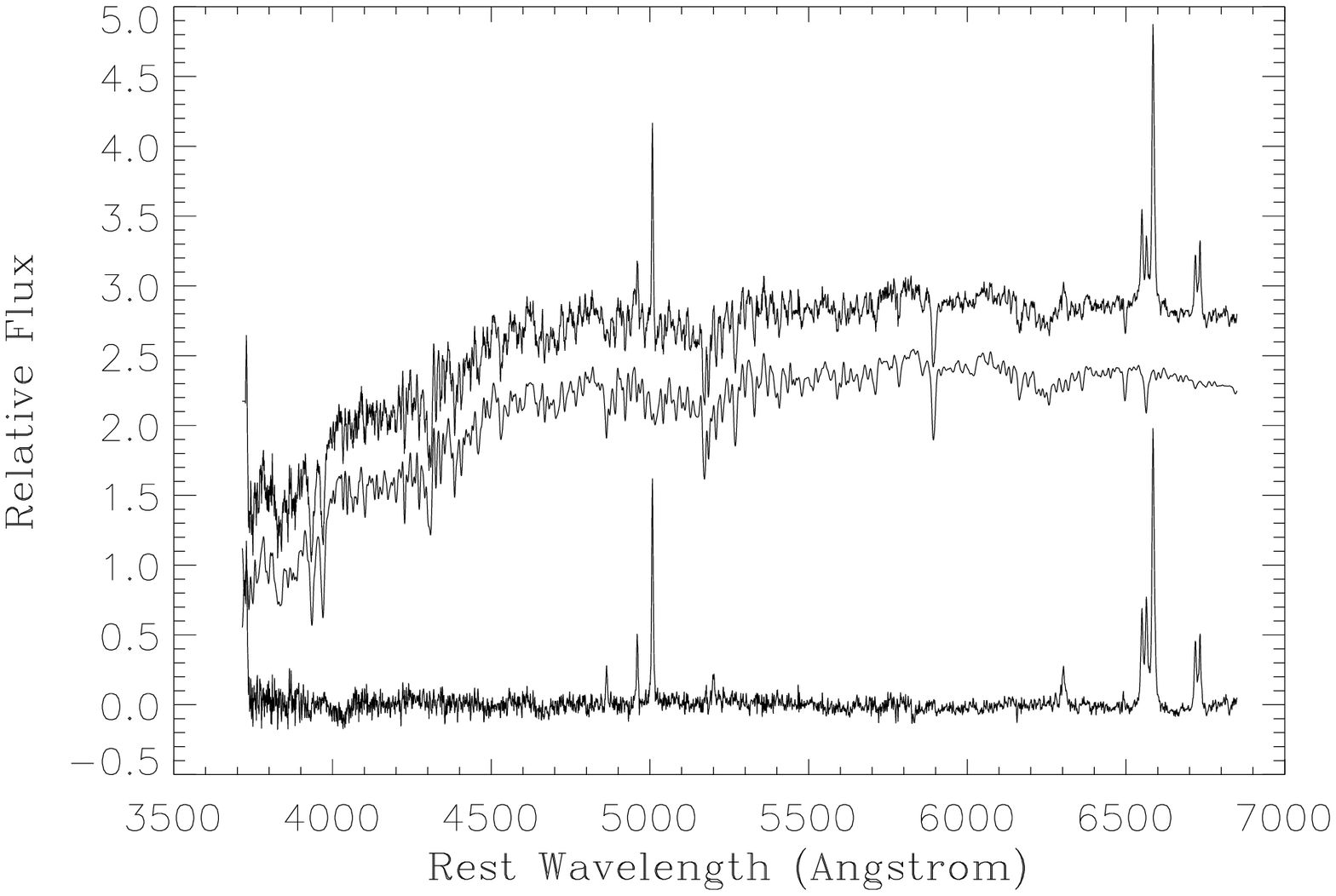} \plotone{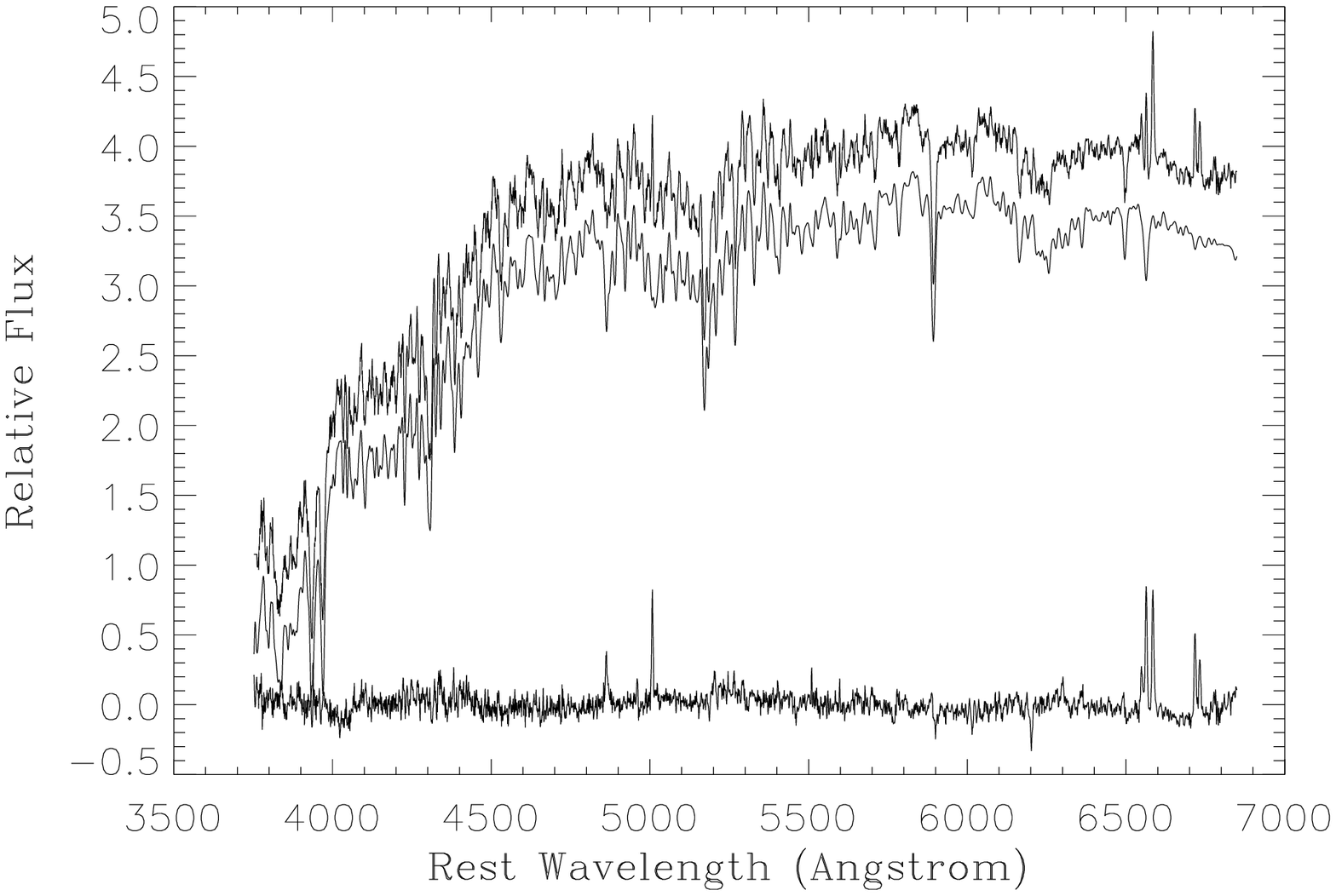}
\plotone{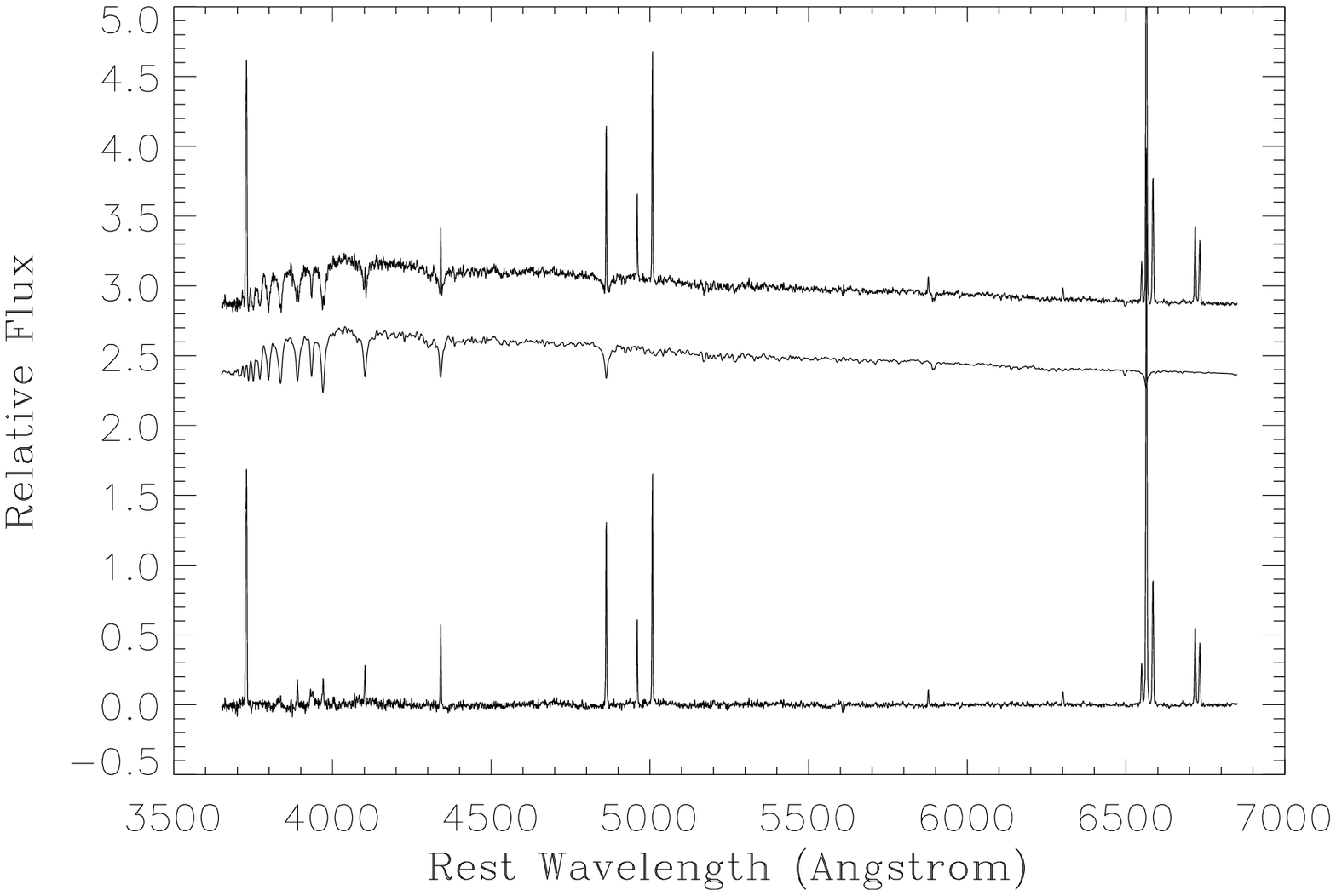} \plotone{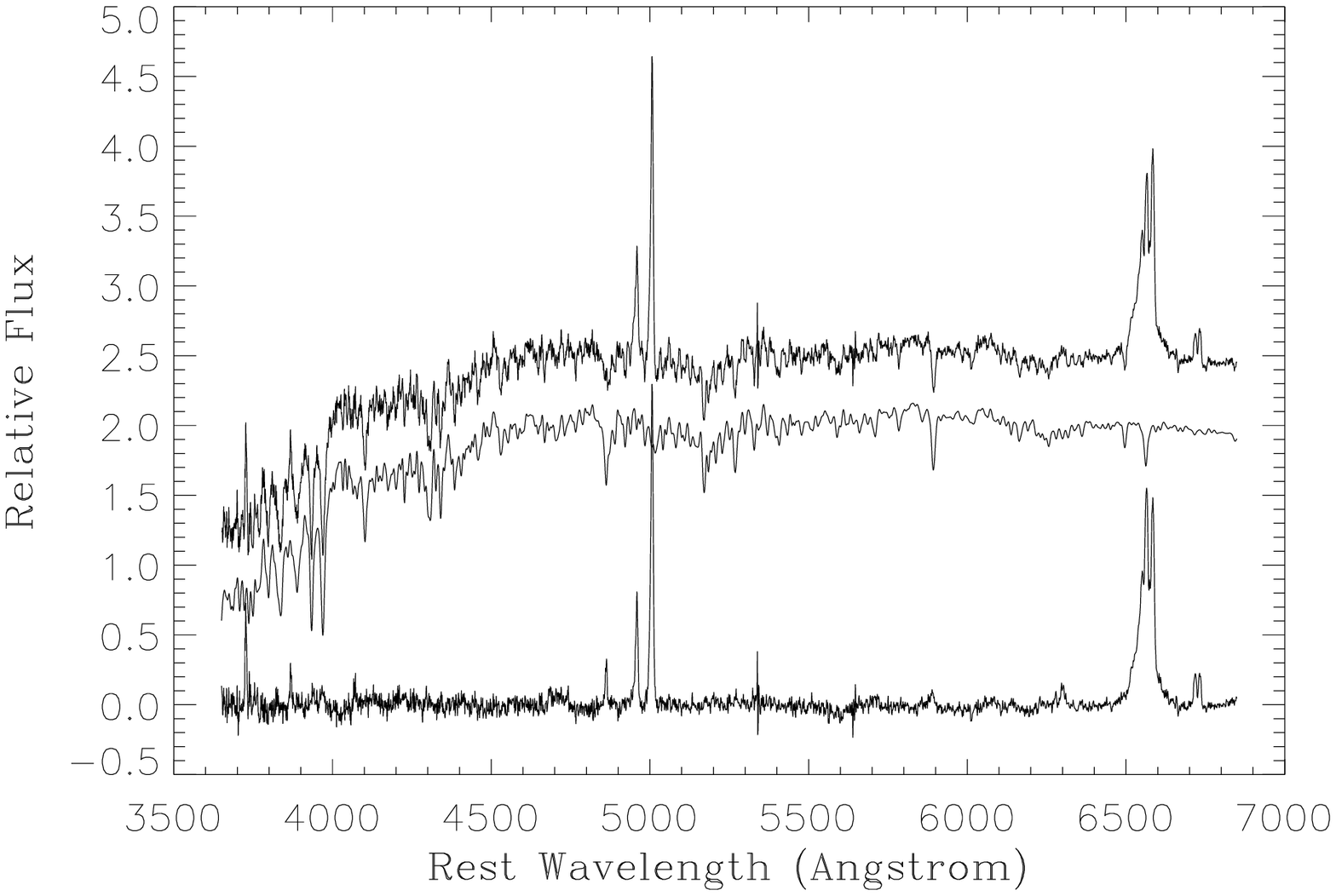}
\plotone{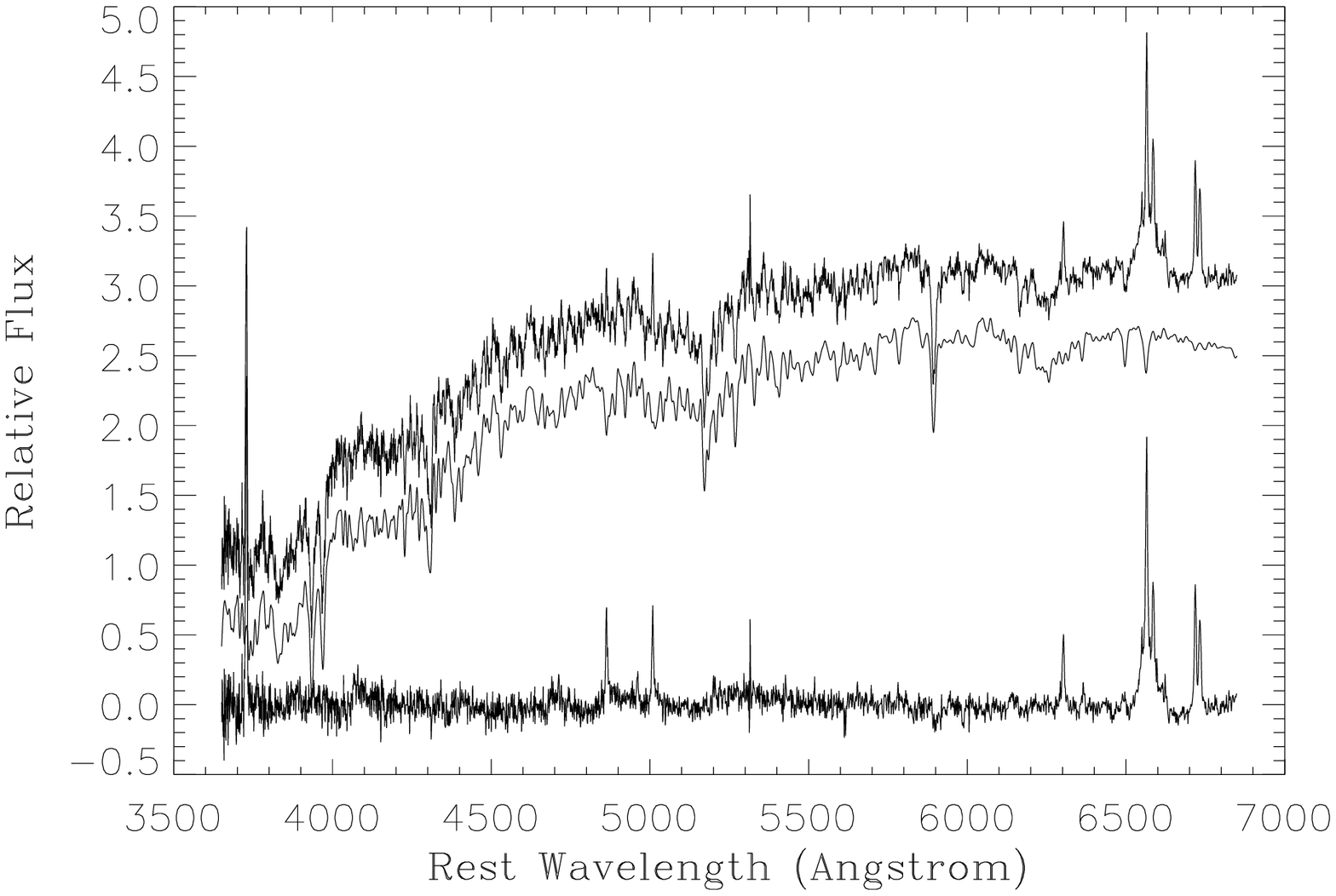} \plotone{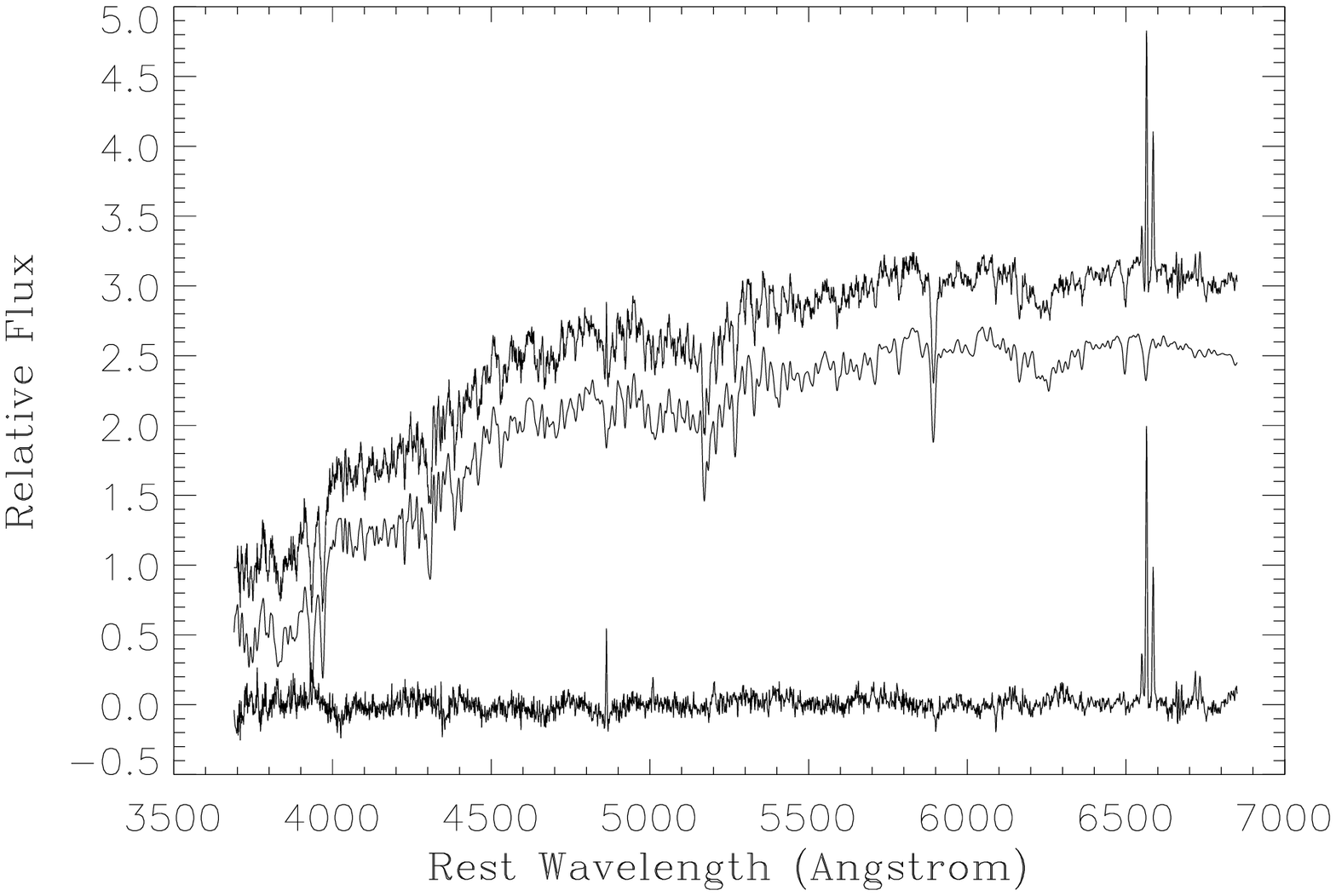}
\plotone{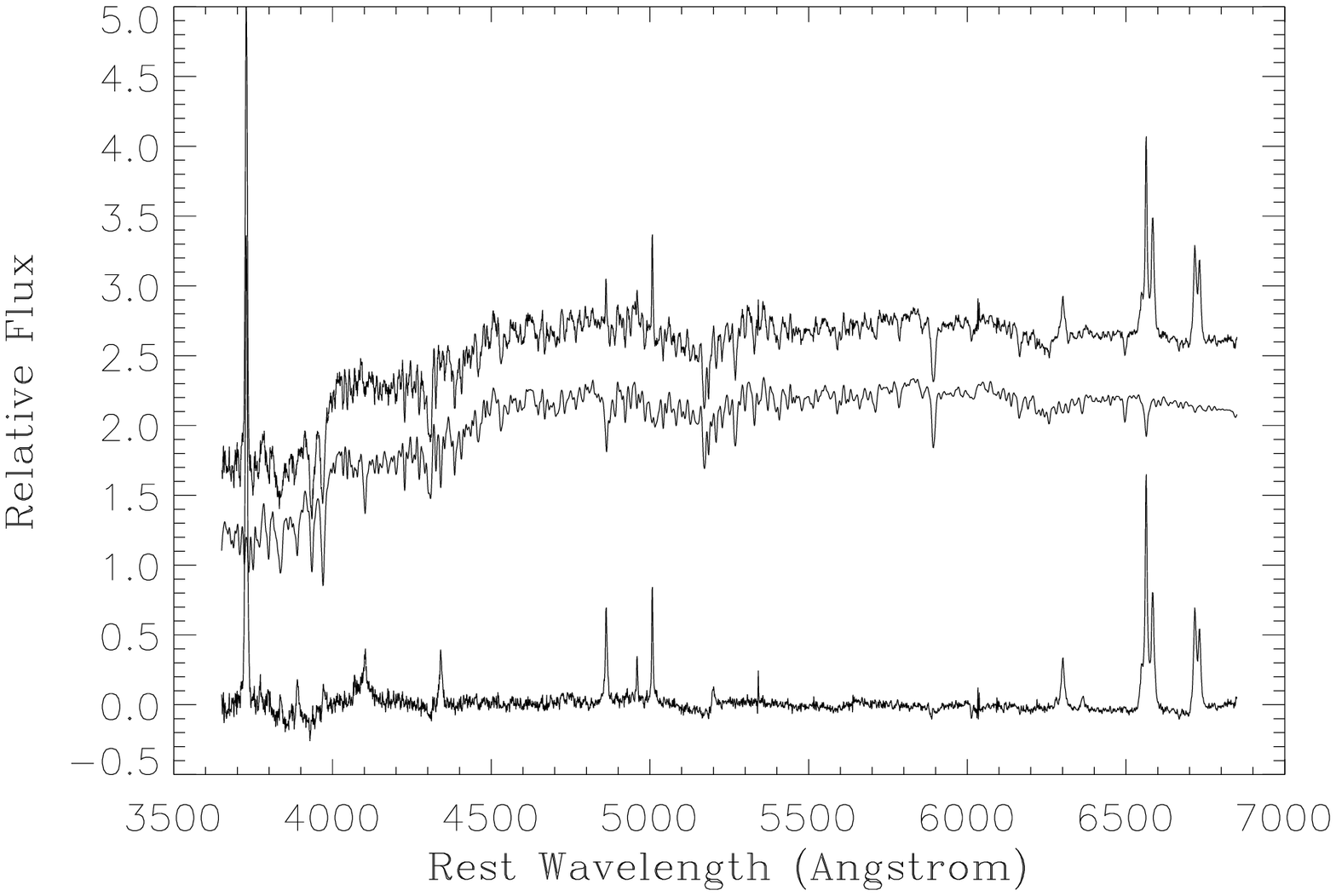} \plotone{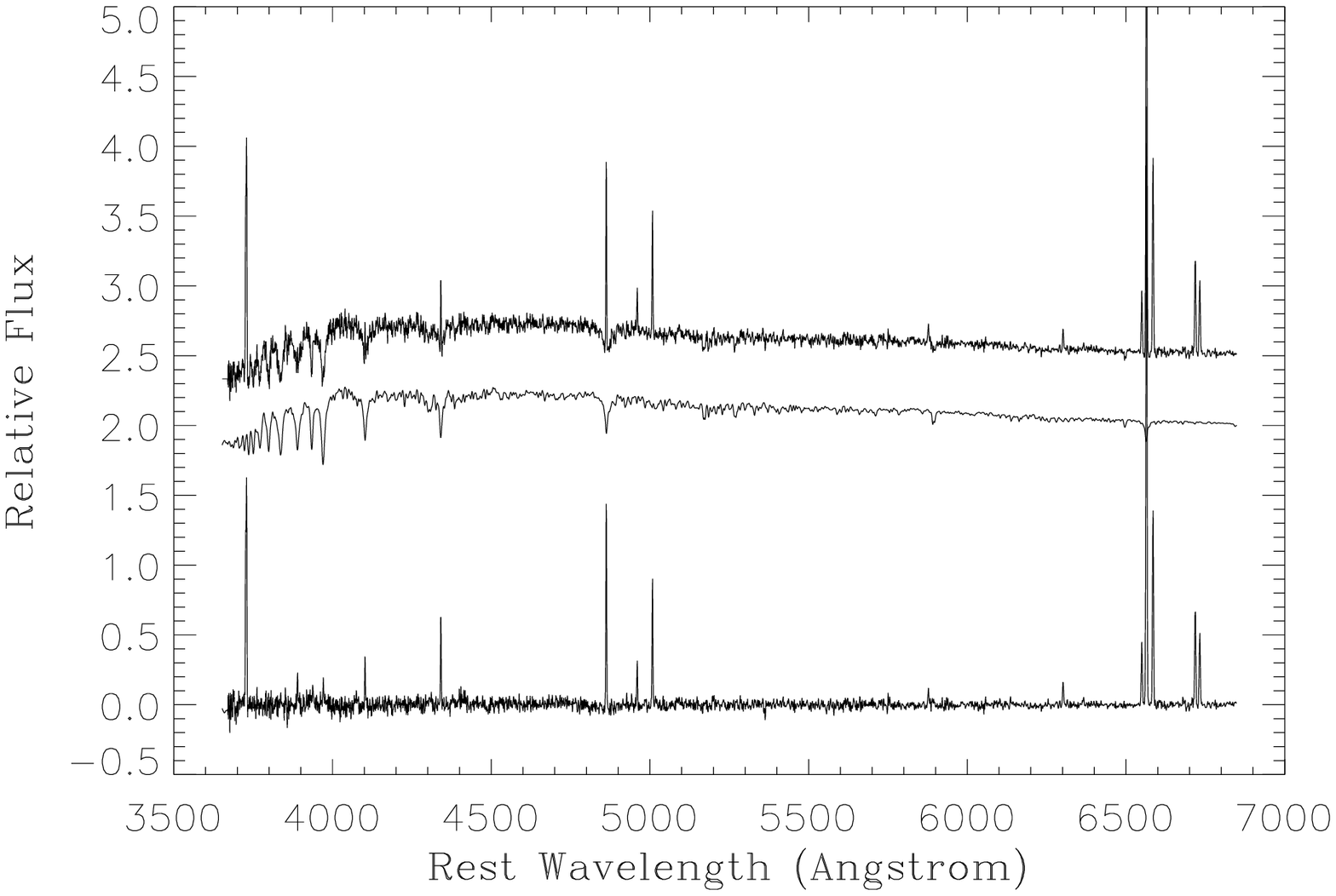}
\caption{
Examples of starlight removal. In each panel,
the three lines (from top to bottom)
are the observed, the modeled and the starlight-subtracted spectra.
For clarity, arbitrary constants are added to the observed and
the modeled spectra.}
\label{f11}
\end{figure}

Fig. \ref{f11} shows the examples of the fits.
Overall residuals in the non-emission line region are consistent with 
flux fluctuations.
The figure also illustrates the importance of the 
stellar modeling to the emission line measurements.
In some of the original spectra, even 
\hbeta is hardly visible, whereas it 
can be easily measured after the underlying starlight subtracted.
Furthermore, the intensities of both \hbeta and \halpha are
modified substantially and their ratio changes to be close
to the theoretical value.

We would like to point out that, 
although most of the galactic spectra
could be well modeled using this method, the fit is not satisfactory 
for a small number of spectra ($\lesssim 0.8\%$),
out of which, most show peculiar molecular absorption bands
and a small portion show obvious Wolf-Rayet feature
around 4600-4750 \AA.
The result is expected because of the lack of very late type stars
and Wolf-Rayet stars in the
stellar library we used (see Fig. \ref{f2}).  Modeling these objects 
requires the improvement of the stellar library.

\begin{figure}
\epsscale{0.6} \plotone{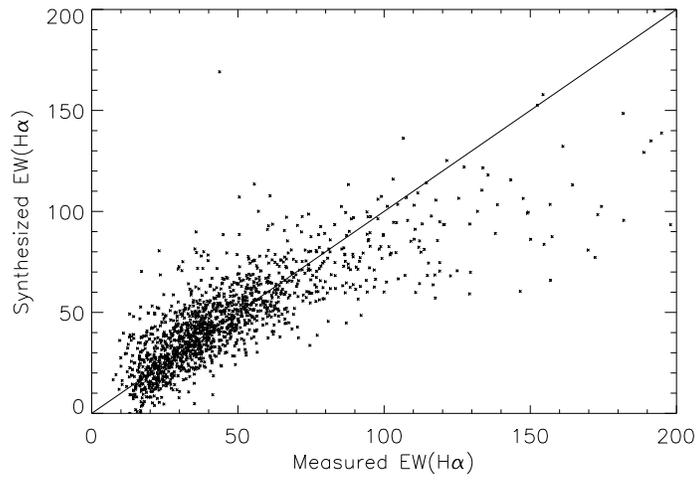}
\caption{A comparison of the \halpha EW as derived by
a Gaussian fit to the \halpha line in the starlight-subtracted spectrum
and by synthesizing the modeling coefficients of templates.}
\label{f12}
\end{figure}

For each spectrum, the best-fitting model is then subtracted from
the original spectrum, yielding a pure emission-line spectrum,
from which emission line parameters are measured.
It is found that the measured equivalent widths of emission lines
are well correlated with the coefficients of the eigen-spectra
obtained during the modeling, though the latter reflects only the
information of the stellar component of galaxies. From SDSS DR1,
we select $\sim$500 spectra that have relatively strong emission
lines, and use the measured equivalent widths of \halpha and the
modeling coefficients to solve a set of linear equations of
\begin{equation}
EW_{H_\alpha}^j = \sum_{i=1}^{9} a_i c_{ij},
\end{equation}
where $EW_{H_\alpha}^j$ is the measured equivalent width of
\halpha of the $jth$ spectrum, and $c_{ij}$ is the expansion coefficient
of the $ith$ eigen-spectrum for the $jth$ spectrum.
Using the resultant constants \{$a_i$\}($i=1,9$),
the \halpha equivalent width of each galaxy can be synthesized
as a linear combination of the modeling coefficients of
the galactic eigen-spectra.
Fig. \ref{f12} presents the relation between the measured
and the synthesized equivalent widths of \halpha.
It can be seen that, for moderately strong emission line,
i.e., $EW(H_\alpha)\lesssim 80$\AA, the synthesized value
is well correlated with the measured one,
while for stronger lines the synthesized values are
smaller than the measured.

\subsection{Stellar Velocity Dispersion}\label{vdisp}

\subsubsection{Tests of the measurement routine}
One of the merits of the method presented in this paper
is that the stellar velocity dispersion of galaxies can be
determined as a byproduct. In order to estimate the accuracy of
the measurement of stellar velocity dispersion, we created a set 
of testing spectra as follows.
First, all the 204 stars in STELIB are classified into seven
groups according to their spectral types, i.e. O, B, A, F, G, K,
and M-type. The average spectra of these groups are then combined
by specifying various fractions, to create a set of 462
spectra. The spectra are then
broadened by convolution with Gaussian of $\sigma$ in the range of 40-300
\kms. Finally, Gaussian noise is added to the broadened spectra to
give S/N = 10-120 per pixel. 

\begin{figure}
\epsscale{0.6} \plotone{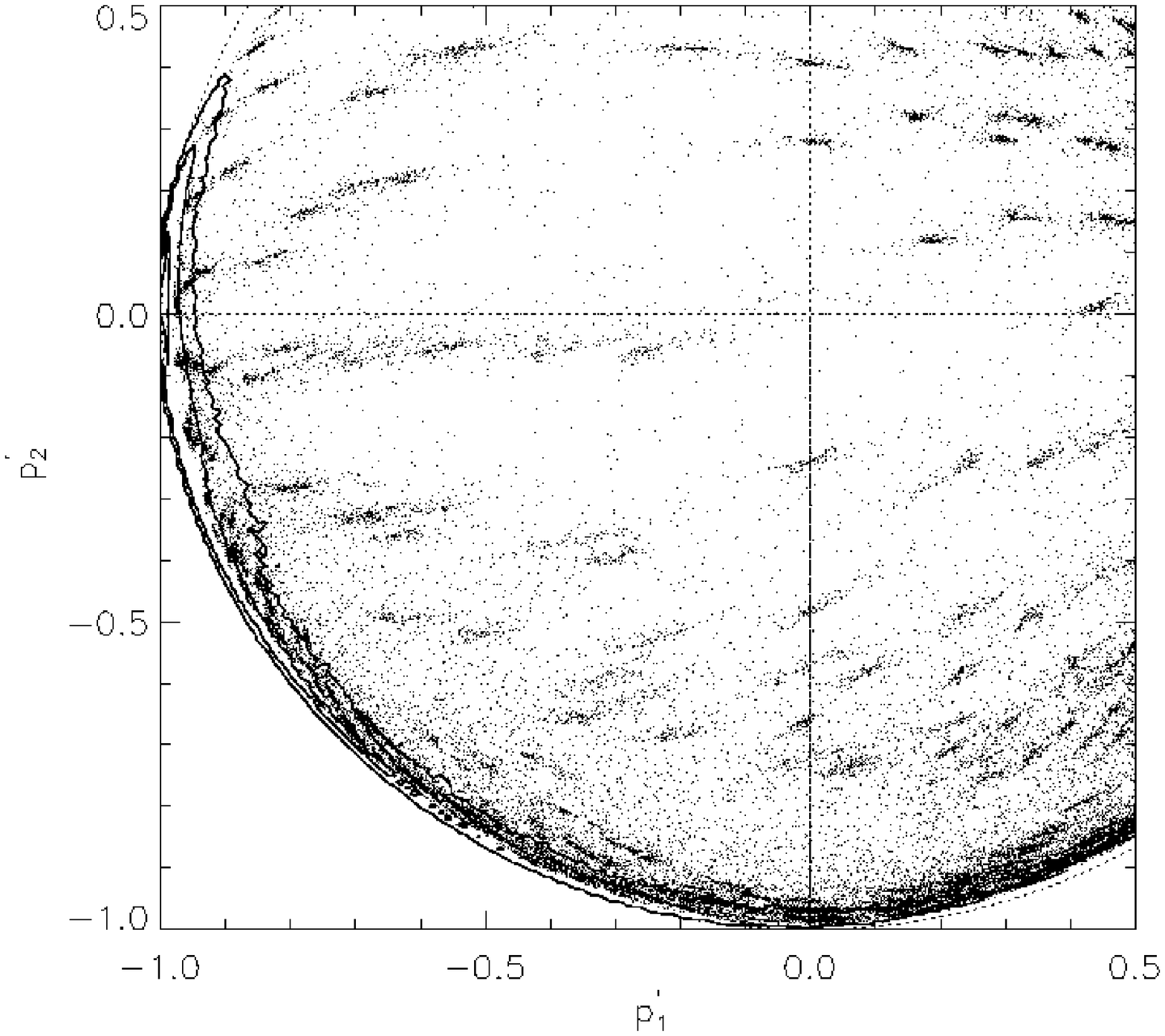} \plotone{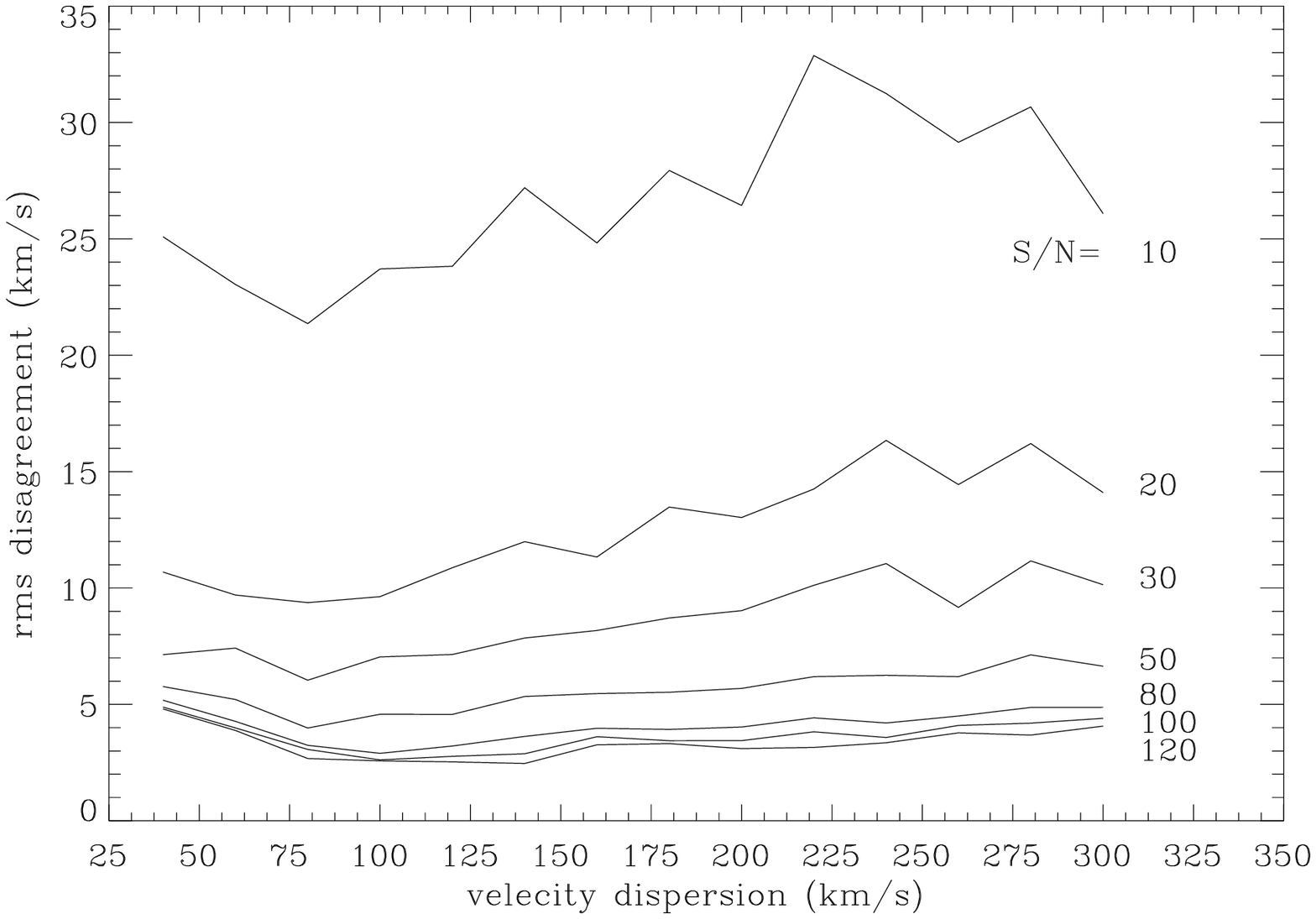}
\caption{
{\it Upper:} Contributions of the first 2 eigen-spectra to the 59136 created
spectra with given stellar velocity dispersion, stellar populations,
signal-to-noise ratios. The three thick lines are the contours
of number density ($n$) of all DR1 galaxies with levels of $\lg(n)=1,2,3$.
{\it Lower:} rms disagreement between the measured and input velocity dispersion
of the 12,820 created spectra with given stellar velocity dispersion, stellar population, and signal-to-noise ratio.}
\label{f13}
\end{figure}

The resultant 59,136 spectra with known velocity dispersion, 
stellar population and S/N ratio are modeled using the method
described above. Fig. \ref{f13} shows the relative contributions of 
the first 2 galactic eigen-spectra to the modeled spectra.
Clearly only a portion of the created spectra (12,820) locate
in the region overlapped with that of SDSS galaxies, and are
selected for later test.

The velocity dispersion for the artificial galaxies is measured in the 
same way as for the really spectra, except that the masked regions for  
emission lines are selected randomly from the real SDSS spectra.  
We find that our method can accurately recover the velocity
dispersion over the entire range of 40-300 \kms. Figure \ref{f13}
shows the root-mean-square (rms) disagreement between the measured
and input velocity dispersion. As a whole, the measured dispersion
is within 35 \kms of the input velocity dispersion for S/N$>$10. 
For synthesized spectra with $S/N\simeq 20$, the typical
value of the SDSS galaxies, the uncertainties of the
velocity dispersion measurements is between $\sim$4\% and 
$\sim$10\% for higher velocity dispersion ($\sigma_*>75$\kms), 
while estimates of low 
velocity dispersions are less accurate ($\sim$20\%). For higher 
signal-to-noise ratios, the measurement errors decrease rapidly. We 
would like to point out that the accuracy of the stellar velocity 
dispersion determined using the method presented in this paper is 
comparable to previous works at high S/N ratio. For the spectra with 
input dispersion range of 50-300 \kms and S/N=120, Barth, Ho \& Sargent 
(2002) yielded an accuracy within 6 \kms , while this value is less than 
5 \kms using our method (see Fig. \ref{f13}). However, our approach should 
over-perform these methods using only a couple of absorption lines at  
low S/N ratio.

\subsubsection{Comparison with the SDSS pipeline}
Stellar velocity dispersion is also measured by the SDSS
spectroscopic pipeline using two different methods:
"Fourier-fitting" and "Direct-fitting"
\footnote{http://www.sdss.org/dr1/algorithms/veldisp.html}. 
The latter is similar to the method presented in this paper
except that
the SDSS pipeline uses only 32 K and G giant stars in M67 as
templates. The SDSS velocity dispersion estimates are obtained by
fitting the rest frame wavelength range 4000-7000 \AA, and then
averaging the estimates provided by the "Fourier-fitting" and
"Direct-fitting" methods. The error on the final value of the
velocity dispersion is determined by adding in quadrature the
errors on the two estimates. The typical error is between
$\Delta(\lg \sigma_*)\sim$ 0.02 dex and 0.06 dex, depending on the
signal-to-noise of the spectra.

\begin{figure}
\epsscale{0.6} \plotone{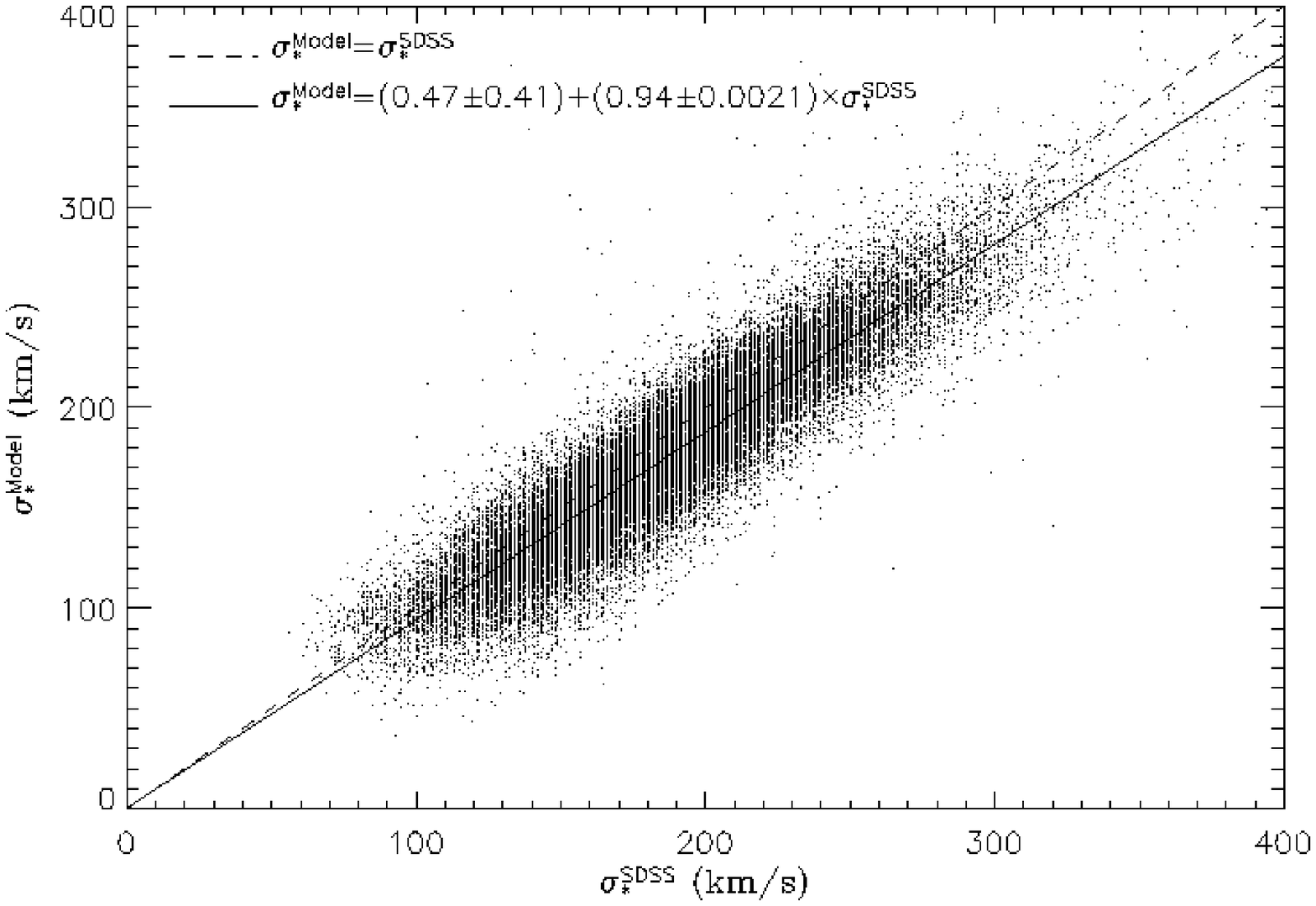} \plotone{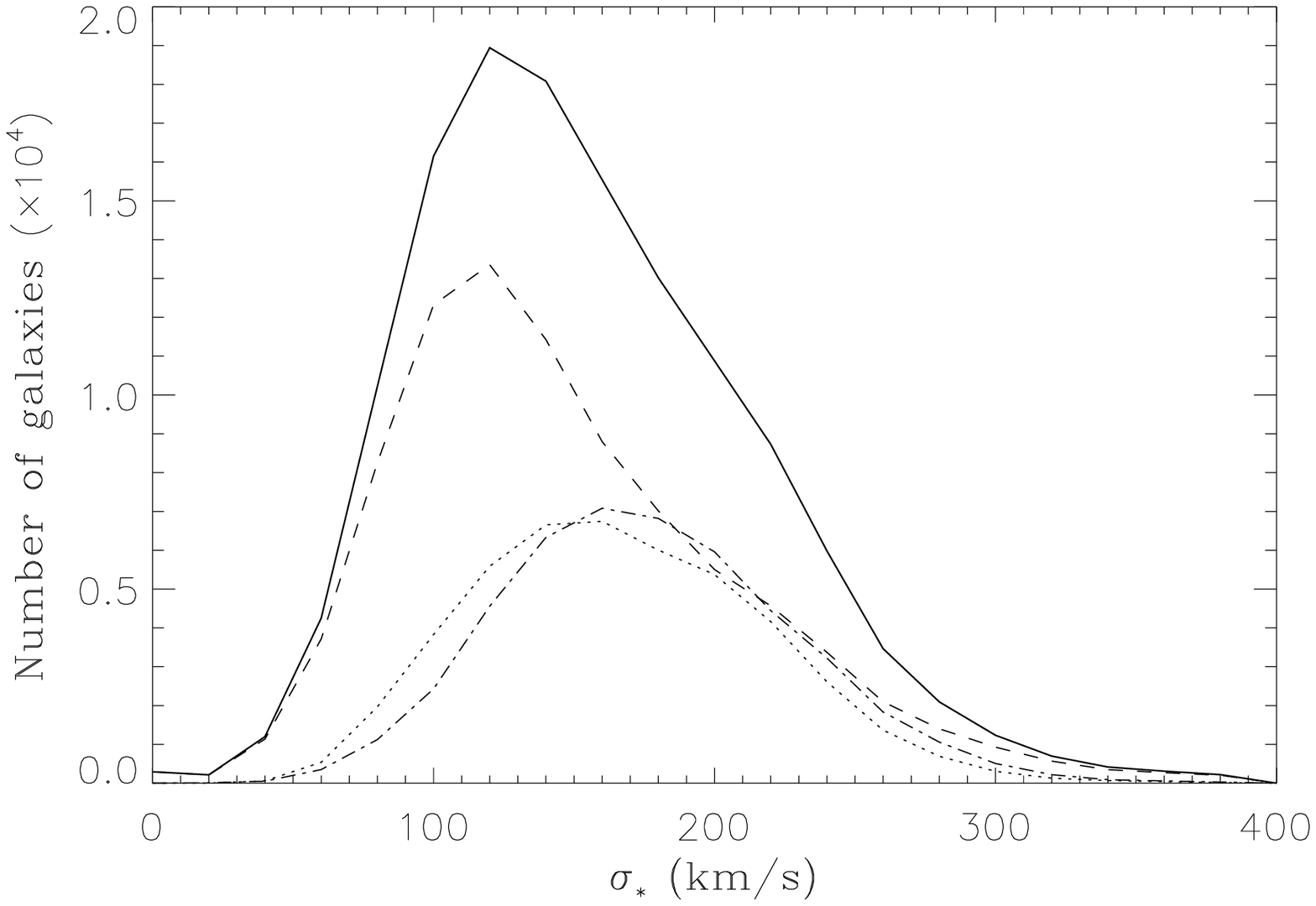}
\caption{
{\it Left}: 
Comparison with stellar velocity dispersion provided by the SDSS pipeline.
The SDSS velocity dispersion $\sigma_*^{SDSS}$ and
that obtained using the present method $\sigma_*^{Model}$ for 46229
spectra in SDSS DR1 are plotted ({\it dots}). The solid line is
the fitted relation and the dashed is that of $\sigma_*^{Model}=\sigma_*^{SDSS}$.
{\it Right}: 
Distribution of stellar velocity dispersion of DR1 
galaxies measured by our method:
{\it dot}: $\sim 4.6\times 10^4$ galaxies that have SDSS
velocity dispersions (the corresponding SDSS velocity dispersions are plotted in
{\it dot-line});
{\it dashed}: the other $\sim 8.8\times 10^4$ galaxies without SDSS
velocity dispersion;
{\it solid}: All the DR1 galaxies.
}
\label{f14}
\end{figure}

However, stellar velocity dispersions $\sigma _{*}$ are only
measured for spheroidal systems by the SDSS spectroscopic
pipeline. The main selection criteria is the PCA classification
"eClass" less than -0.02, typical of spectra of early-type galaxies.
Adopting these criteria, $\sim 4.6\times 10^4$ spectra of galaxies in
SDSS DR1 were chosen to measure $\sigma _{*}$ by the SDSS
pipeline. The values determined by the SDSS pipeline are compared
with our measurements in Fig. \ref{f14} (left panel).
The two estimates
are almost consistent with each other except that our measurements
are systemically a bit smaller than those determined by the SDSS
pipeline. This is understandable because our model is expected to fit 
the spectrum better, reducing the template mismatch problem. The 
iterative rejection of emission lines in our procedure can also improve 
the accuracy of the velocity measurement.

In Fig. \ref{f14} (right panel), the distribution of the
velocity dispersion of this paper and that provided by
the SDSS pipeline are presented. In addition to the $\sim 4.6\times
10^4$ galaxies, we also measure the rest $\sim 8.8\times 10^4$
galaxies whose $\sigma _{*}$ are not provided by the SDSS
pipeline. It can be seen that the velocity dispersions of these
objects are systematically smaller than that of the $\sim 4.6\times
10^4$ early type-galaxies. This is consistent with the general
impression that the stellar velocity dispersion of spheroidal
systems is generally larger than that of late-type ones. Using the
method presented in this paper, the data set of $\sigma _{*}$ is
enlarged $\sim 3 $ times compared with that provided by the SDSS
pipeline. As it is illustrated in Fig \ref{f6}, by carefully
masking the emission-line wavelength range, $\sigma _{*}$ can be
reliably determined using our method for most of emission line galaxies.

\subsection{Spectral Classification}
Galaxy classification plays important role in the study of galaxy
formation and evolution. Three methods are often used for this
purpose: morphological segregation, rest-frame colors, and direct
spectrum-based classifications, while each method has its own
unique drawbacks and advantages (Madgwick et al. 2003).

As shown in Fig. \ref{f4},
the position angle $\theta$ of the stars on the diagram of 
their relative contributions to the first two
eigen-stars are well correlated with the spectral type,
suggesting that $\theta$ is a good indicator of stellar types
and thus useful for stellar classification.
Similarly, most of the galaxies on the diagram of 
their relative contributions to the first two eigen-galaxies 
also distribute on the unit circle (see Fig. \ref{f8}).
The corresponding position angle of galaxies
is therefore expected to be a good single-parameter
classifier of galaxies.
However, such a PCA-based classification of galactic spectra
is only applicable to small samples, 
such as the galaxy sample presented in \S\ref{samsel},
because of the required computation expense of
implementing PCA on large or very high-dimensional data sets.
Nevertheless, we note that there is another similar parameter,
namely the position angle $\theta^\prime$ of galaxies
on the diagram of the contributions of
the first two eigen-galaxies to the modeled spectra
(the $p_1^\prime~ vs~ p_2^\prime$ plane; see Fig. \ref{f13}),
which is also a good classifier.

\begin{figure}
\epsscale{0.6} \plotone{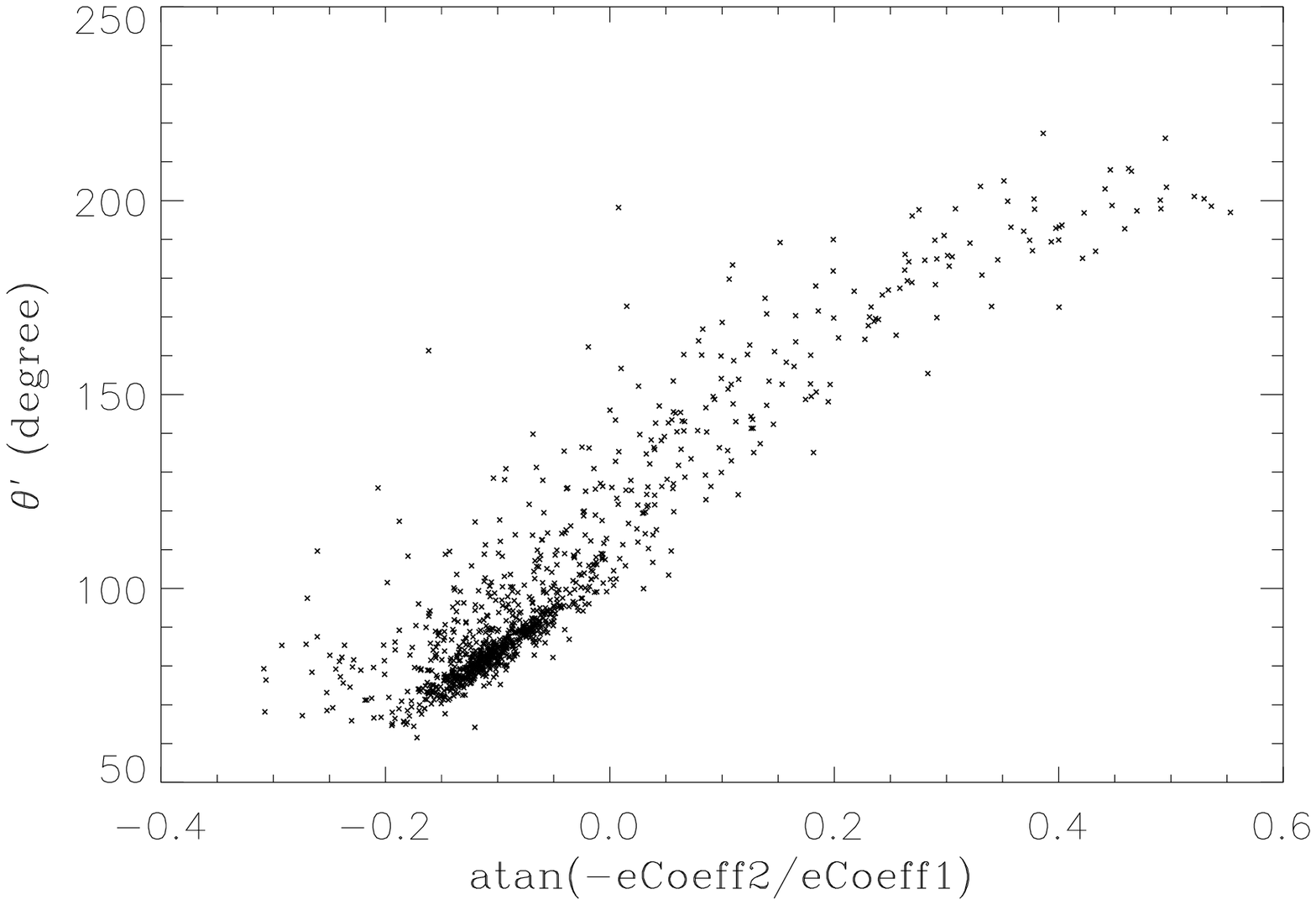} \plotone{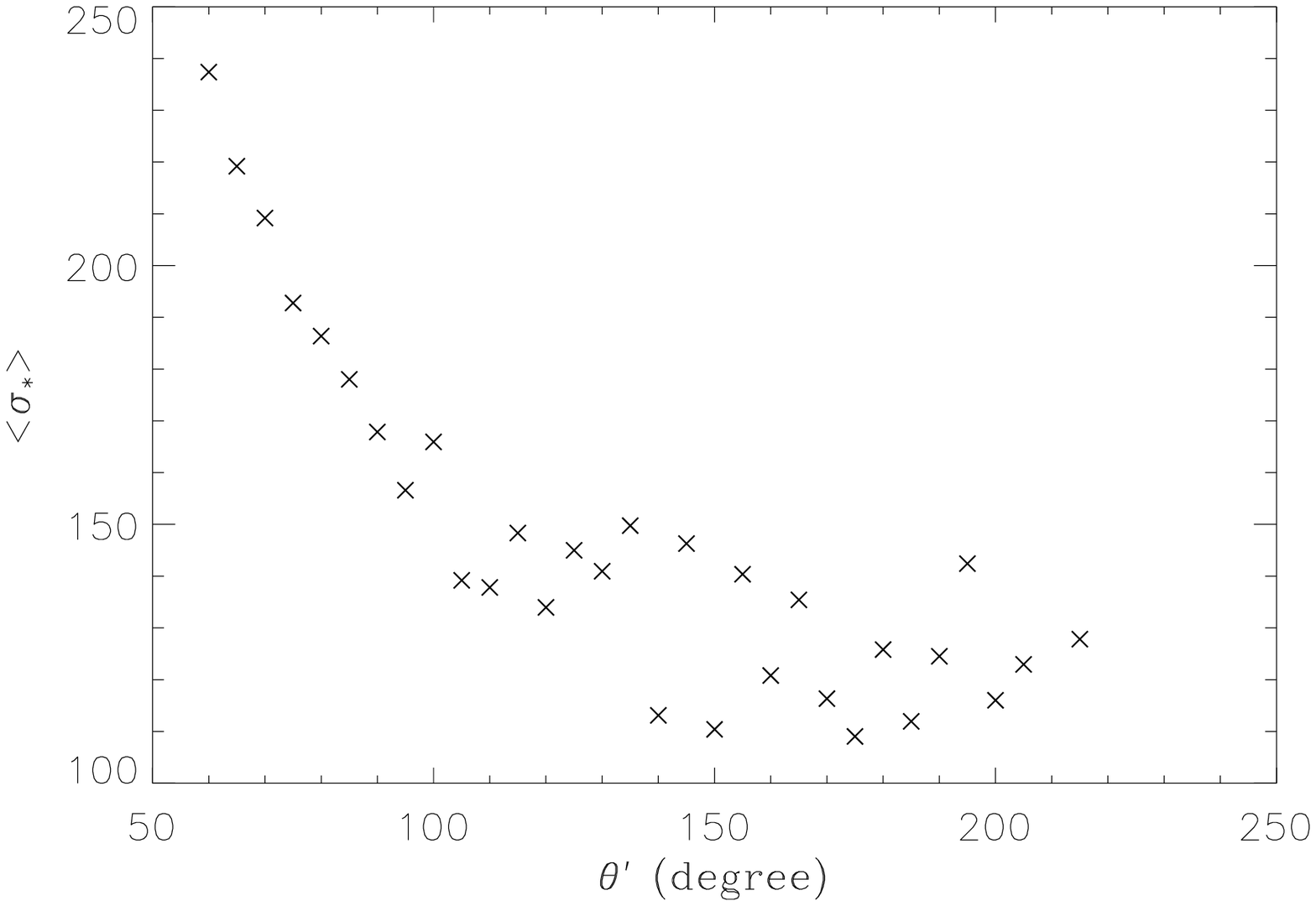}
\caption{
{\it Left}: The position angles $\theta^\prime$
of the 1000 galaxies on the
$p_1^\prime$ {\it vs}\rm\ $p_2^\prime$ diagram
(left panel in Fig. \ref{f13}; 
$\theta^\prime$ is measured anticlockwise from the positive Y-axis)
versus their atan(eCoeff2/eCoeff1) provided by the SDSS pipeline.
{\it Right}: Mean velocity dispersion of 1000 galaxies as a function
of their position angles on the $p_1^\prime$ {\it vs} $p_2^\prime$ diagram.
}
\label{f15}
\end{figure}

In fact, SDSS DR1 pipeline also provide a spectral classification of 
galaxies, by cross-correlating with eigen-templates constructed from 
early SDSS spectroscopic data using PCA method. 5 eigen-coefficients 
and a classification number are stored in parameters "eCoeff1-5" and 
"eClass", respectively (Stoughton, et al. 2002). The parameter "eClass", 
based on the expansion coefficients "eCoeff1"-"eCoeff5", ranges from 
about -0.35 to 0.5 for
early- to late-type galaxies. Since the sign of the second
eigen-spectrum has been reversed with respect to that of the Early
Data Release (EDR, see Stoughton et al. 2002), the expression
$atan(-eCoeff2/eCoeff1)$ is recommended rather than "eClass" as
the single-parameter classifier
\footnote{http://www.sdss.org/dr1/algorithms/redshift\_type.html}.
To compare our classifier $\theta^\prime$ with that provided by
the SDSS pipeline, we randomly select 1000 galaxies from SDSS DR1
and fit their spectra to obtain the modeling coefficients of
templates and subsequently the position angles $\theta^\prime$.
As may be seen clearly in Fig. \ref{f15} (left panel), the position angle
$\theta^\prime$ is actually a good single-parameter classifier,
which is well correlated with the
classifier $atan(-eCoeff2/eCoeff1)$ provided by the SDSS pipeline.
This is understandable because both of the classifiers are obtained
based on PCA. Moreover, our classifier $\theta^\prime$
is also well correlated with the mean velocity dispersion
$<\sigma_*>$ of the 1000 galaxies (see the right panel in Fig. \ref{f15}),
with smaller $\theta^\prime$ corresponding to larger velocity dispersion,
thus to earlier type galaxies.

\begin{figure}
\epsscale{0.6} \plotone{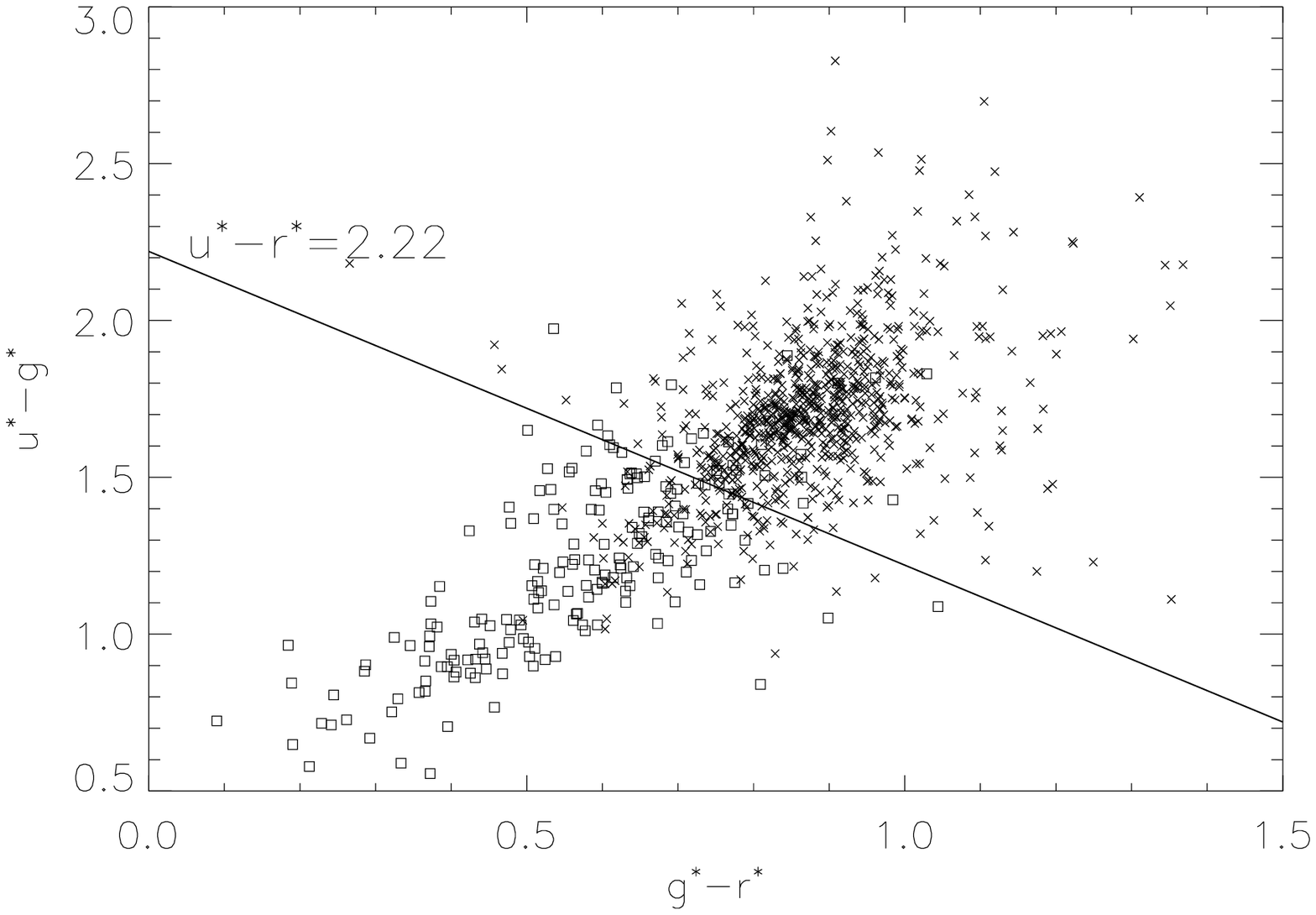} \plotone{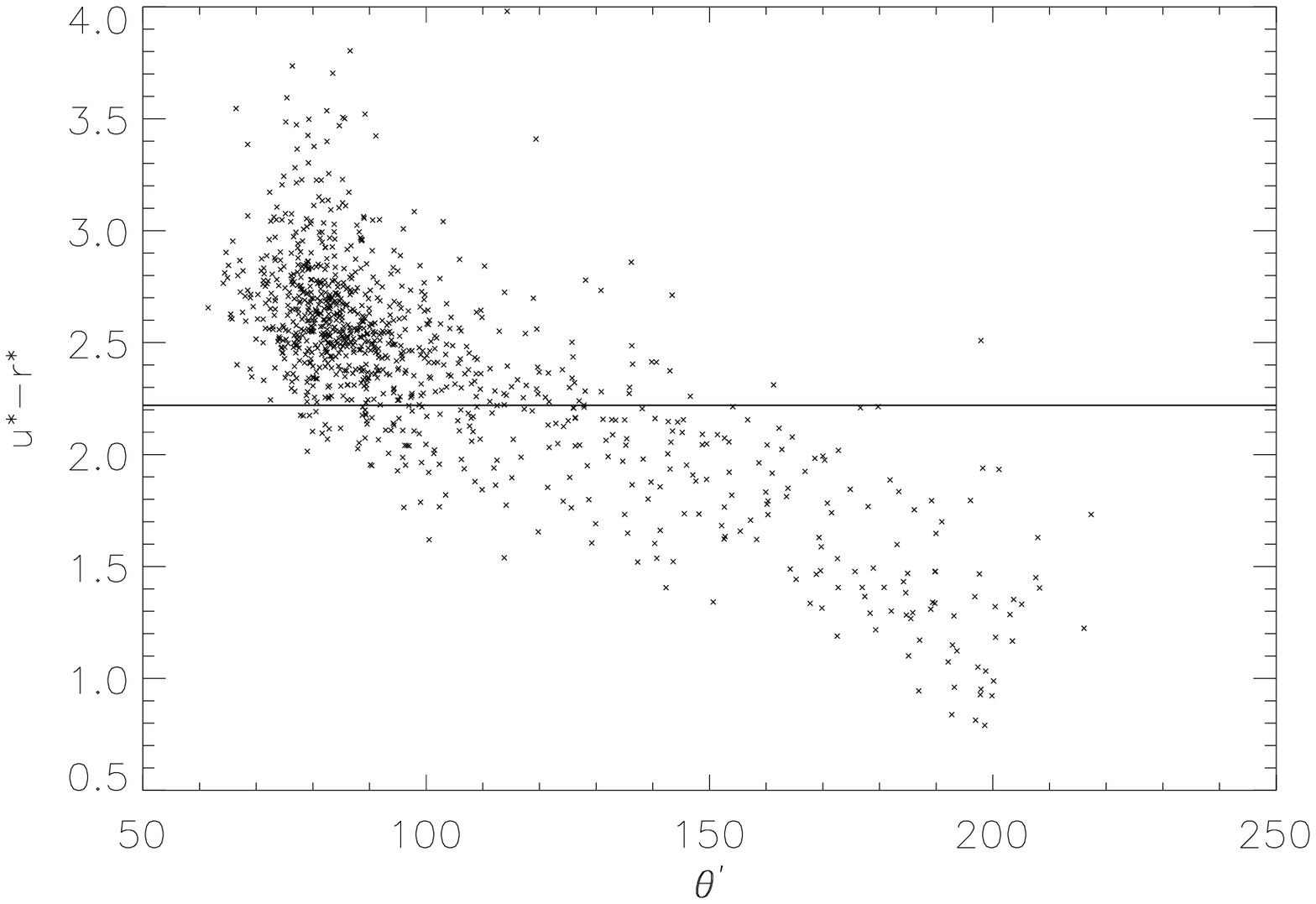}
\caption{
{\it Left}: Color-Color diagram for 1000 galaxies.
{\it cross}\rm : galaxies with position angle $\theta^\prime > 120^\circ$;
{\it square}\rm : galaxies with $\theta^\prime < 120^\circ$ (see Fig. \ref{f13}).
The solid line denotes $u^*-r^*=2.22$ .
{\it Right}: $u^*-r^*$ {\it vs}\rm\ $\theta^\prime$ (see Fig. \ref{f13}) for
the 1000 galaxies. The horizontal line denotes $u^*-r^*=2.22$.
}
\label{f16}
\end{figure}

Using the rest-frame colors, Strateva et al. (2001) studied the
color distribution of a large uniform sample of galaxies detected
in SDSS commissioning data and showed that the $g^*-r^*$ versus
$u^*-g^*$ color-color diagram is strongly bimodal, with an optimal
color separation of $u^*-r^*=2.22$ and the two peaks corresponding
roughly to early- (E, S0, and Sa) and late-type (Sb, Sc, and Irr)
galaxies. 
As can be seen clearly in 
Fig. \ref{f16}, the early- and late-type galaxies separated by
$u^*-r^*=2.22$ could be also well classified using our
single-parameter classifier $\theta^\prime$,
indicating that $\theta^\prime$ might be also compatible
for galactic classification. 

\subsection{Near-infrared SED}

Once a spectrum has been modeled using the 9 galactic
eigen-spectra, the contributions of the 24 eigen-stars to
the modeled spectrum can be conveniently obtained.
Since the 24 stellar eigen-spectra contain the infrared information
of the 204 original stars (\S3), the infared fluxes at J,H and K$_{s}$ bands
could be re-constructed from a set of expansion coefficients

We test reliability of the stellar SED in the near infrared using this 
method on a set of "average" spectra of normal galaxies along the Hubble 
diagram between E and Sc between 0.1 and 2.4 $\mu$m, presented by 
Mannucci et al. (2001) \footnote{http://www.arcetri.astro.it/~filippo/spectra}.
The composited spectra were derived from 28 nearby galaxies, and
overall precision of the calibrated spectra is about 2 per cent.

\begin{figure}
\epsscale{0.6} \plotone{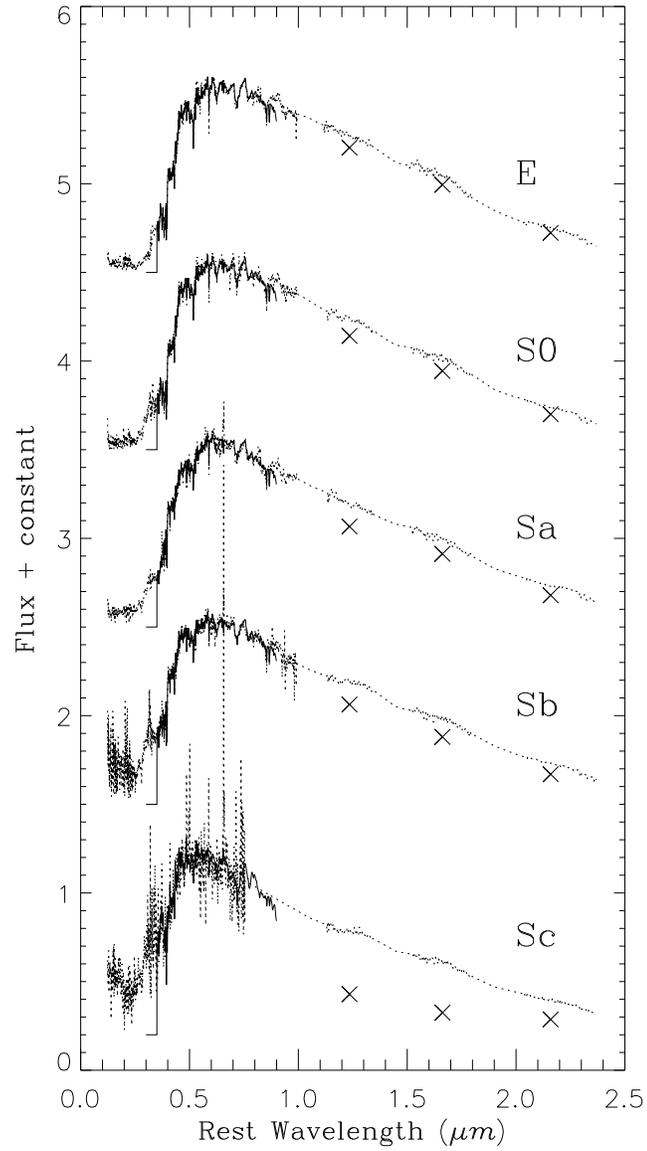}
\caption{
Modeling the "average" spectra of normal
galaxies along the Hubble diagram between E and Sc.
\it{dashed lines}\rm: the "average" spectra.
\it{solid lines}\rm: the modeled optical spectra.
\it{cross}\rm: the modeled J, H and K fluxes.
Arbitrary constants were added to the scaled spectra
for clarity.
}
\label{f17}
\end{figure}

The five spectra are modeled using the 9 templates and the
results are presented in Figure \ref{f17}. As can be seen, the
modeled infrared fluxes are systematically lower than the observed 
one by $\sim 10\%$ for E to Sb types, and 30\%
for Sc galaxies. The discrepancy is understandable as the observed 
infrared flux includes additional emission from hot dust, presumably 
caused by stochastic heating of small grains by UV photons, which is 
expected to increase both with the dust content and young stellar 
population along the Hubble sequence. Besides, if dust extinction is 
patched, the single E(B-V) correction will under-estimate the true 
extinction in the optical.  
Nevertheless, we argue that the result of our
method could give reliable information of starlight at infrared
band for a galaxy (especially for early types),
by only using its optical spectrum.

\subsection{Average Reddening}

\begin{figure}
\epsscale{0.6} \plotone{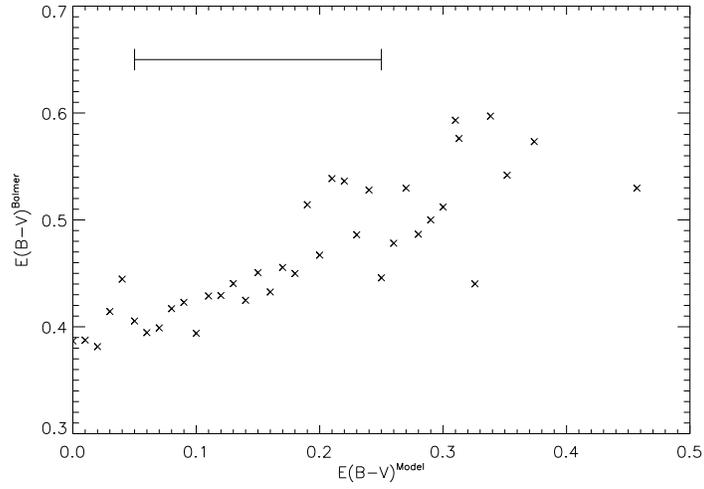}
\caption{Mean color excess $E(B-V)^{Balmer}$ estimated
from the flux ratio of Balmer lines $f(H_\alpha)/f(H_\beta)$ {\it vs}
the modeled $E(B-V)^{Model}$ obtained using the method presented in this paper,
for $\sim 10^4$ H{\sc ii} regions or starburst galaxies in SDSS DR1.
The bin size of the color excess is so chosen as to have
at least 30 galaxies in each bin.
The horizontal line in the upper-left corner is the typical error of $E(B-V)^{Model}$.
}
\label{f18}
\end{figure}

Our solution also yields an average stellar reddening $E(B-V)^*$ for each 
galaxy (see \S4.2). The extinction is quite significant for 85.5\% galaxies, 
with the $E(B-V)^*$ in the range of 0.01-0.5.  On the other hand, the internal 
reddening can be also estimated using Balmer decrements of the emission 
lines from H{\sc ii} region. To check the consistency of the two estimates,
we select a sample of $\sim 10^4$ H{\sc ii} galaxies from SDSS DR1 
adopting the criteria of Kewley et al. (2001):
\begin{equation}
\log{\big{(}\frac{[\mbox{O {\sc iii}}]\lambda 5007}{\mbox{H}\beta}\big{)}} <
\frac{0.61}{\log([\mbox{N {\sc ii}}]/\mbox{H}\alpha)} + 1.19.
\end{equation}
Using the effective extinction curve
$\tau_\lambda=\tau_V(\lambda/5500$\AA$)^{-0.7}$,
which was introduced by Charlot \& Fall (2000),
the color excess arising from attenuation by dust in the selected
galaxy $E(B-V)^{Balmer}$, can be written:
\begin{equation}
E(B-V)^{Balmer} = A_V/R_V = 1.086\tau_V/R_V,
\end{equation}
\begin{equation}
\tau_V=-\frac{\ln[F(\mbox{H}\alpha)/F(\mbox{H}\beta)]
-\ln[I(\mbox{H}\alpha)/I(\mbox{H}\beta)]}
{(\lambda_{\mbox{H}\alpha}/5500)^{-0.7}-(\lambda_{\mbox{H}\beta}/5500)^{-0.7}},
\end{equation}
where $I(\mbox{H}\alpha)/I(\mbox{H}\beta)$=2.87 is the intrinsic
Blamer flux ratio , $F(\mbox{H}\alpha)/F(\mbox{H}\beta)$ the
observed Balmer flux ratio
, $\tau_V$ the effective V-band optical depth,
$\lambda_{\mbox{H}\alpha}$=6563\AA,
$\lambda_{\mbox{H}\beta}$=4861\AA, and $R_V$=3.1. 
For objects with the observed flux
ratio $F(\mbox{H}\alpha)/F(\mbox{H}\beta)<$ 2.87, $\tau_V$ is set to
zero. The resultant $E(B-V)^{Balmer}$ {\it vs} $E(B-V)^{*}$ is plotted in 
Fig. \ref{f18}.
Although the dispersion of
$E(B-V)^{*}$ is large ($\sim$ 0$^m$.2), the tendency is
apparent: the $E(B-V)^{*}$ increases with increasing
$E(B-V)^{Balmer}$, and the former is systemically smaller than the
latter, which is also consistent with previous works 
(e.g Calzetti, Kinney, \& Storchi-Bergmann 1994).

\section{Summary}

We have developed an empirical method for modeling the stellar spectrum 
of galaxies. The absorption-line templates with zero velocity dispersion 
are constructed based on two successive applications of Principal Component 
Analysis (PCA), first to 204 stars in the stellar library STELIB, then 
to a uniform sample of galaxies selected from SDSS DR1.
With 9 templates, we can fit quite well the stellar spectra of galaxies 
in the SDSS DR1. As byproducts, the stellar velocity dispersion,
the near-infrared SED, the spectral type, and the average reddening
are determined simultaneously. For a spectrum of S/N=20, typical for 
SDSS galaxies, the velocity dispersion can be determined to accuracy 
of $\sim$4-10\% at larger value ($>$75~km~s$^{-1}$), and $\sim$20\% at 
low dispersion ($<$75~~km~s$^{-1}$).  The average reddening of stellar 
light is correlated, and is systematically smaller than that derived 
from Balmer decrements of H{\sc ii} region. After stellar light have been 
subtracted, emission line parameters are measured for all galaxies. An 
interesting result is that the measured equivalent width of H$\alpha$ is well 
correlated with the modeling coefficients of the templates, suggesting 
that these coefficients also contain information of current star formation 
rate. 

The success of this approach could be understood from the following aspects. 
First, two applications of PCA highly concentrate the most prominent features
from both the stellar library and the galaxy sample into the first few 
galactic eigen-spectra, and hence significantly reduce the number of the 
final templates. This makes the procedure of spectral modeling much quicker,
and the results much more stable and reliable, in particular, when dealing with 
large datasets, in comparison with using direct stellar populations. 
Second, the templates derived from the large galaxy sample with full 
coverage of spectral type, ensure a close match to the true underlying 
stellar population of different type galaxies, including emission line 
galaxies that contain much young stellar component. 
As a result, the modeled spectra reflect the true information
on the stellar population in galaxies, and provide useful
clues for their spectral classifications.
Third, 
the templates are not obtained directly from the observed galactic spectra,
but from the modeled one with zero velocity dispersion,
and therefore could match better the absorption line profile 
of galaxies.
Finally, the extended wavelength coverage
(from optical to near-infrared) in the templates enables us
to study the stellar component at NIR band, using only
the optical spectrum of galaxies.

It should be pointed out that there are still rooms for improvement of this 
method.  First, lack of very late type stars limits our application of 
templates in modeling a small fraction of galaxies which show prominent, 
peculiar molecular absorption features. 
This can be improved by adding very late 
type stars to the stellar library. Second, implication of each PCs should 
be tackled, by setting a relation between fitting coefficients of templates 
and stellar population. This may be addressed by combining the stellar 
population synthesis with the PCA method, which is still under 
investigation. 

\acknowledgements

We thank Dr. T. Kiang for help in improving the English.
This work is supported by the Chinese National Science foundation
through NSF 10233030, BaiRen project of Chinese Academy of Science and 
a key program of Chinese ministry of science and technology.
This work is also partly supported by the Excellent Young Teachers
Program of MOE, P.R.C. This paper has made use of the stellar
spectra libraries STELIB and HILIB, and the data from 2MASS and
SDSS.

Funding for the creation and distribution of the SDSS Archive has
been provided by the Alfred P. Sloan Foundation, the Participating
Institutions, the National Aeronautics and Space Administration,
the National Science Foundation, the U.S. Department of Energy,
the Japanese Monbukagakusho, and the Max Planck Society. The SDSS
Web site is http://www.sdss.org/.

The SDSS is managed by the Astrophysical Research Consortium (ARC)
for the Participating Institutions. The Participating Institutions
are The University of Chicago, Fermilab, the Institute for
Advanced Study, the Japan Participation Group, The Johns Hopkins
University, Los Alamos National Laboratory, the
Max-Planck-Institute for Astronomy (MPIA), the
Max-Planck-Institute for Astrophysics (MPA), New Mexico State
University, Princeton University, the United States Naval
Observatory, and the University of Washington.


\end{document}